\documentclass[preprint]{aastex}
\usepackage{graphicx}
\usepackage{natbib}
\newcommand{\ms}{$\,$M$_\mathrm{\odot}$}
\newcommand{\ls}{$\,$L$_\mathrm{\odot}$}

\newcommand{\be}{\begin{equation}}
\newcommand{\ee}{\end{equation}}

\newcommand{\el}[2]{\ensuremath{^{#1}\mathrm{#2}}}

\newcommand{\pg}{\ensuremath{\mathrm{(p,\gamma)}}}

\newcommand{\an}{\ensuremath{\mathrm{(\alpha,n)}}}

\shorttitle{3D hydro simulations of a PIE}
\shortauthors{R.~J. Stancliffe et al.}

\title{3D hydrodynamical simulations of a proton ingestion episode in a low-metallicity asymptotic giant branch star}
\author{Richard J. Stancliffe\altaffilmark{1,2}, David S.~P. Dearborn\altaffilmark{3}, John C. Lattanzio\altaffilmark{1}, Stuart A. Heap\altaffilmark{1} and Simon W. Campbell\altaffilmark{1}}

\altaffiltext{1}{Monash Centre for Astrophysics (formerly the Centre for Stellar and Planetary Astrophysics), Monash University, VIC 3800, Australia}
\altaffiltext{2}{Research School of Astronomy and Astrophysics, Mount Stromlo Observatory, Cotter Road, Weston Creek, ACT 2611, Australia}
\altaffiltext{3}{Lawrence Livermore National Laboratory, 7000 East Avenue, Livermore, CA 94551, U.S.A.}
\email{rjs@mso.anu.edu.au}
\begin{document}

\begin{abstract}
We use the 3D stellar structure code {\sc djehuty} to model the ingestion of protons into the intershell convection zone of a 1\ms\ asymptotic giant branch star of metallicity $Z=10^{-4}$. We have run two simulations: a low resolution one of around 300,000 zones, and a high resolution one consisting of 2,000,000 zones. Both simulations have been evolved for about 4 hours of stellar time.  We observe the existence of fast, downward flowing plumes that are able to transport hydrogen into close proximity to the helium burning shell before burning takes place. The intershell in the 3D model is richer in protons than the 1D model by several orders of magnitude and so we obtain substantially higher hydrogen-burning luminosities -- over $10^8$\ls\ in the high resolution simulation -- than are found in the 1D model. Convective velocities in these simulations are over 10 times greater than the predictions of mixing length theory, though the 3D simulations have greater energy generation due to the enhanced hydrogen burning. We find no evidence of the convective zone splitting into two, though this could be as a result of insufficient spatial resolution or because the models have not been evolved for long enough. We suggest that the 1D mixing length theory and particularly the use of a diffusion algorithm for mixing do not give an accurate picture of these events. An advective mixing scheme may give a better representation of the transport processes seen in the 3D models.
\end{abstract}


\section{Introduction}

The final stage of the life of a low-mass star (one in the mass range 1-6\ms) is known as the asymptotic giant branch (AGB) phase. During this part of its life, the star consists of an inert core of carbon and oxygen, surrounded by two burning shells with  the interior one burning helium and the outer one burning hydrogen. This double shell burning configuration is unstable and there are periodic episodes of runaway helium burning. These are referred to as thermal pulses \citep[see the reviews by][for more details]{1983ARA&A..21..271I,2005ARA&A..43..435H}. During these thermal pulses, helium burning drives a convective region between these helium- and hydrogen-burning shells. The advance of this intershell convection region is checked by the increase in entropy across the hydrogen burning shell.

As one moves to stars of lower metallicity, the entropy barrier provided by the hydrogen burning shell is weakened. If the metallicity is low enough, the intershell convection zone can penetrate up into the hydrogen-rich regions \citep{1996ApJ...459..298C, 2000ApJ...529L..25F}. Protons are then drawn deep into the star where they burn vigorously. This is a proton ingestion episode (PIE)\footnote{Proton ingestion can also occur during the core helium flash for models of very low metallicity \citep[see e.g.][]{1990ApJ...349..580F, 1990ApJ...351..245H}.}. The occurrence of these PIEs in one-dimensional stellar evolution codes has recently been studied by \citet{2008A&A...490..769C}, \citet{2009MNRAS.396.1046L},  \citet{2009PASA...26..139C}, \citet{2009PASA...26..145I} and \citet{2010MNRAS.405..177S}. In these codes, if a sufficient quantity of protons are ingested, the energy released when these burn can lead to the intershell convection zone splitting into two parts. The lower part remains driven by helium burning, while the upper part is now sustained by hydrogen burning. While all authors seem to agree that some splitting of the convection zone takes place, there is little agreement on the mass and metallicity regimes where this can occur.

One must take the results of one dimensional evolution codes with a degree of scepticism: they are dependent on our treatment of convection. Convection, as implemented in the form of mixing length theory \citep{1958ZA.....46..108B}, has long been the Achilles' Heel of stellar evolution computations and there are several reasons to be wary of it here. First, convection is inherently a three-dimensional process. Are we missing some important property of the way protons are ingested by reducing it to one dimension? Secondly, mixing is commonly treated as a diffusive process in stellar evolution codes, yet convection is advective in nature. Finally, mixing length theory does not take account of the effect of nuclear burning in downflowing material. Energy released due to nuclear burning will increase the buoyancy of the downward moving element, reducing its speed \citep[see][for further details]{2001ApJ...554L..71H}. For these reasons, it is fruitful to look at the issue of proton ingestion using hydrodynamical modelling.

Hydrodynamical modelling has advanced to the point where we can study convective processes in some detail. This has been done for a wide variety of situations. \citet{2006ApJ...642.1057H} investigated convection in thermal pulses in two dimensions. \citet{2009A&A...501..659M} simulated the core helium flash in low-mass stars in two and three dimension. Carbon and oxygen shell burning in the late stages of the evolution of massive stars have been tackled by \citet{2006ApJ...637L..53M, 2007ApJ...667..448M}. \citet{2006PhDT........20M} and \citet{2011ApJ...733...78A} have also modelled, in two-dimensions, a 23\ms\ star with simultaneously active C, O, Ne and Si burning shells. They contrast this 2D model with 1D stellar evolution calculations, highlighting the fact that convection is advective in nature, rather than diffusive, and that in the case of vigourous convection and burning the diffusion approximation shows large shortcomings. 

Hydrogen ingestion at the core helium flash has been modelled using the hydrodynamic code {\sc herakles} by \citet{2010A&A...520A.114M}. Their simulation was based upon a 1D stellar evolution model for a 0.85\ms\ star of zero metallicity taken from \citet{2008A&A...490..769C}. As an input model, they took the stellar structure at a time when the convective region had already split into two. They then modelled a wedge of this star in both two and three dimensions. They found that convection was not sustained for very long in both the zones. However, there was some uncertainty as to whether this was a numerical artifact brought about by insufficient resolution, or by problems with the initial stabilized input model. Further {\sc herakles} simulations of proton ingestion at the core helium flash (albeit at solar metallicity) by \citet{2011...A&A...XX.XXM} suggest that substantial hydrogen ingestion can indeed create two distinct convective regions, with the upper one being driven by hydrogen burning. These authors note that the existing entropy barrier is permeable to chemical transport when multidimensional flows are considered and thus hydrogen can end up in the helium-driven convective region.

Proton ingestion during a very late thermal pulse\footnote{This is a thermal pulse that takes place when the star's envelope has been all but removed and the star is effectively a white dwarf.} was modelled using 3D hydrodynamic simulations by \citet{2011ApJ...727...89H}. These simulations dealt only with the gas dynamics and no nuclear burning was included. Their computations were performed on Cartesian grids of $576^3$ and $384^3$ zones and consisted of three polytropic layers which reflect the stellar structure. An artificial luminosity was added in a shell to represent the driving of the convection zone by nuclear burning. They evolve their simulation for around 7 convective turnover times, where the turnover time is estimated to be around 3000 seconds. Their simulations suggest that large upwelling convective cells dominate the convective flow patterns and that ingestion of hydrogen takes place in downflows that form when these large cells meet each other. It is suggested that Kelvin-Helmholtz instabilities are the likely main mechanism for the entrainment of protons at the convective-radiative boundary.

For this study, we use the {\sc djehuty} code, which has been developed at Lawrence Livermore National Laboratory and which operates in a parallel environment, to model stars in three dimensions. The major advantage of {\sc djehuty} is that it allows us to model an entire sphere covering the region of interest (rather than just a wedge of the star) and it also includes a comprehensive suite of nuclear burning reactions. The code is described in extensive detail in \citet{2003ASPC..293....1B},  \citet{2005ApJ...630..309D} and \citet{2006ApJ...639..405D}. Here we shall simply recap the salient details.

The mesh that {\sc djehuty} uses to compute a star consists of seven distinct sections, with each section being made up of hexahedral cells of various shapes. At the centre, the mesh has a cube consisting of $N\times N\times N$ cells. To each of the faces of this cube are attached ``arm'' segments consisting of $N\times N\times L$ cells. One of the $N\times N$ faces of each segment is attached point by point to the central cube, while the outermost $N\times N$ face is mapped to lie on a spherical surface. Thus as one proceeds outward through the arm segment, the cells morph from cuboidal structure to wedge shapes so that the mesh transitions from planar to spherical geometry. The number of cells over which this transition takes place can be set by the user. The $N\times L$ faces of adjacent arm segments are also joined in a point by point fashion.
 
The code has a nuclear network consisting of 21 species: \el{1}{H}, \el{3}{He}, \el{4}{He}, \el{12}{C}, \el{13}{C}, \el{13}{N}, \el{14}{N}, \el{15}{N}, \el{15}{O}, \el{16}{O}, \el{17}{O}, \el{18}{O}, \el{17}{F}, \el{18}{F}, \el{19}{F}, \el{20}{Ne}, \el{22}{Ne}, \el{24}{Mg}, \el{28}{Si}, \el{32}{S} and \el{56}{Ni}. The network allows {\sc djehuty} to be employed on a variety of astrophysical probelms including: the core helium flash in low-mass stars \citep{2006ApJ...639..405D}, extra mixing on the giant branch \citep{2006Sci...314.1580E}, the explosion of white dwarfs and supernovae \citep{2005ApJ...630..309D, 2005NuPhA.758..467M}. For the purposes of this work, the network contains all the species we require.

In this paper, we apply {\sc djehuty} to the problem of proton ingestion during thermal pulses in low-mass AGB stars. In section~\ref{sec:models} we describe how we set up the simulations. Section~\ref{sec:results} describes the results of these simulations, which we then compare to the results of one-dimensional stellar evolution calculations in section~\ref{sec:1d}.
 
\section{The Models}\label{sec:models}

To obtain a starting model to use as an input for the 3D runs, we use the 1D stellar evolution code {\sc stars}, orginally developed by \citet{1971MNRAS.151..351E} and updated by many authors \citep[e.g.][]{1995MNRAS.274..964P,2009MNRAS.396.1699S}. We evolve a 1\ms\ star of metallicity $Z=10^{-4}$ from the pre-main sequence to the asymptotic giant branch. We use 999 meshpoints, a mixing length parameter of $\alpha=2.0$ and no convective overshooting is employed. On the AGB, we utilised the AGB-specific modifications outlined in \citet{2004MNRAS.352..984S}. Mass loss is included using the Reimers' prescription \citep{1975MSRSL...8..369R} with $\eta=0.4$ up to the TP-AGB, and the prescription of \citet{1993ApJ...413..641V} is used for the duration of the TP-AGB.

\begin{figure}
\includegraphics[angle=270,width=0.85\columnwidth]{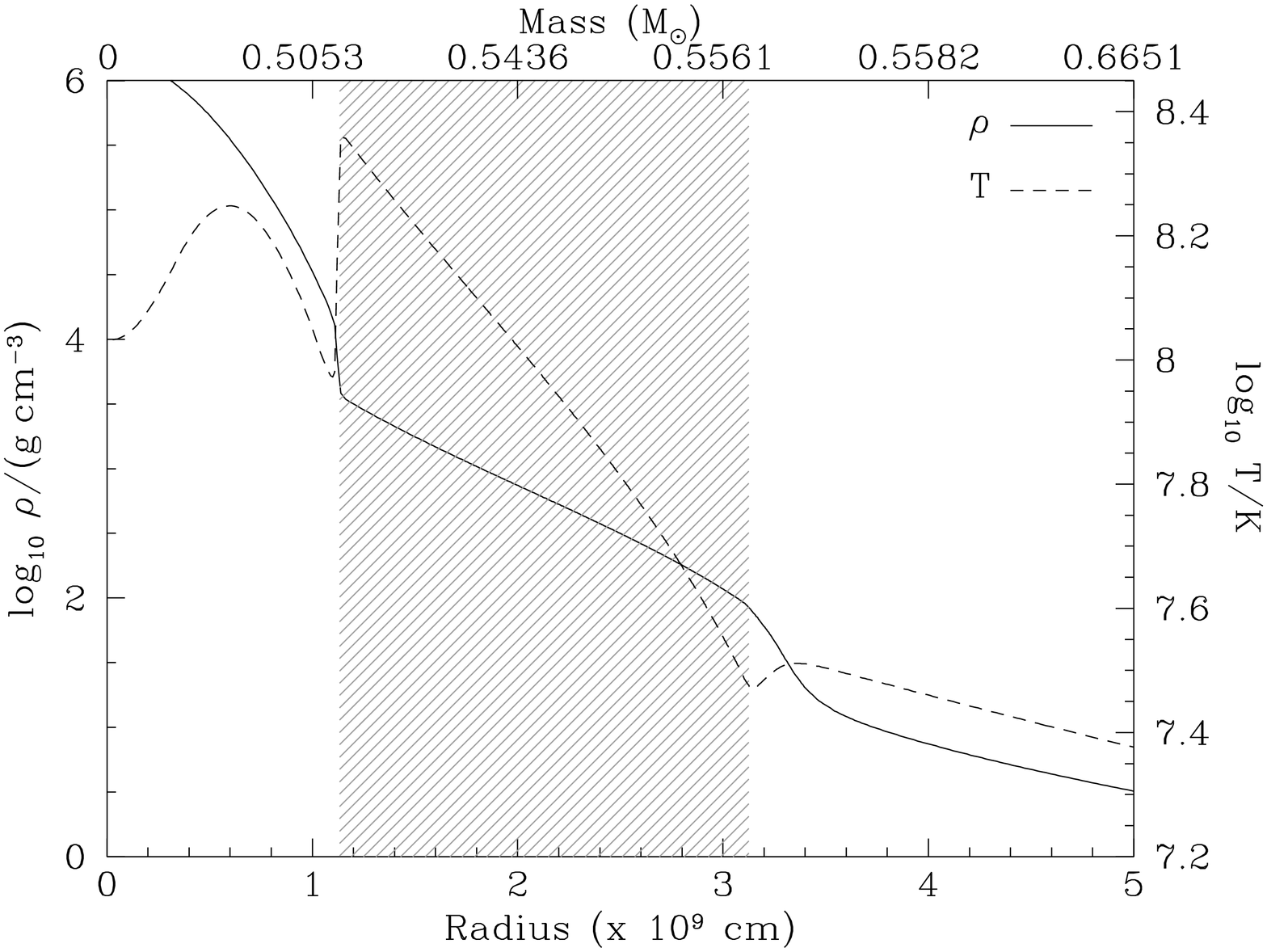}
\includegraphics[angle=270,width=0.81\columnwidth]{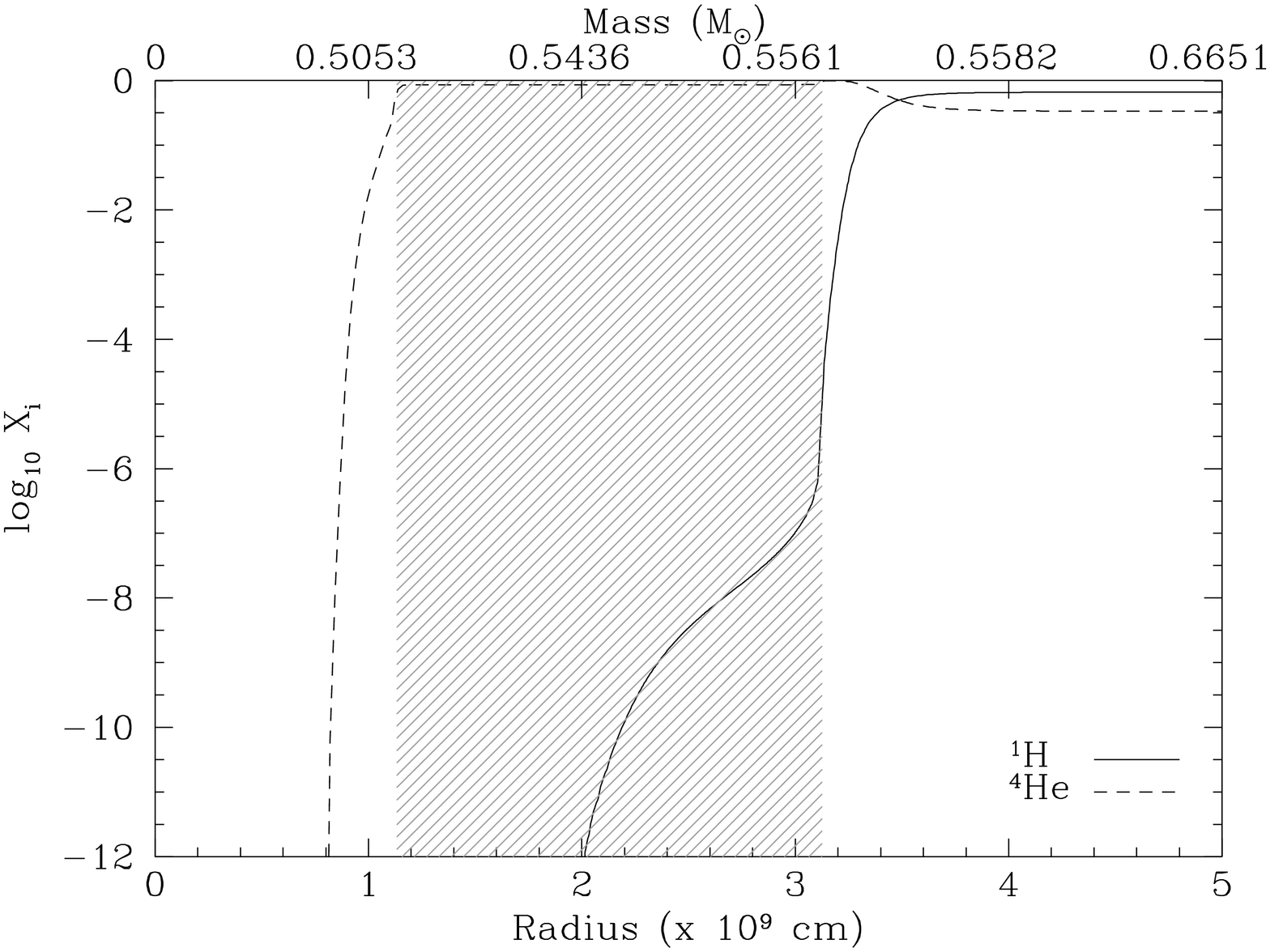}
\caption{{\bf Upper panel:} Density and temperature profiles as a function of radius for our 1D model. {\bf Lower panel:} Abundance profiles of hydrogen and helium-4 as a function of radius for the same model. In both panels, the grey shaded region denotes the extent of the convective zone.}
\label{fig:1Dabunds}
\end{figure}

The model is evolved until just after the peak of the second thermal pulse. As the helium luminosity declines from a peak value of $\log_{10} L_\mathrm{He}/\mathrm{L_\odot} = 6.63$, the intershell convection zone begins to move outward in mass, ingesting protons up to an abundance of around $X_\mathrm{H} \approx 10^{-5}$. It is at this point that we extract a model to use as the input for the 3D runs. The temperature, density and abundance profiles of the 1D model are shown in Fig.~\ref{fig:1Dabunds}. Based on the convective velocity as derived from mixing length theory (MLT), the turnover time for the intershell is approximately one hour.

\begin{figure*}
\includegraphics[width=0.55\columnwidth]{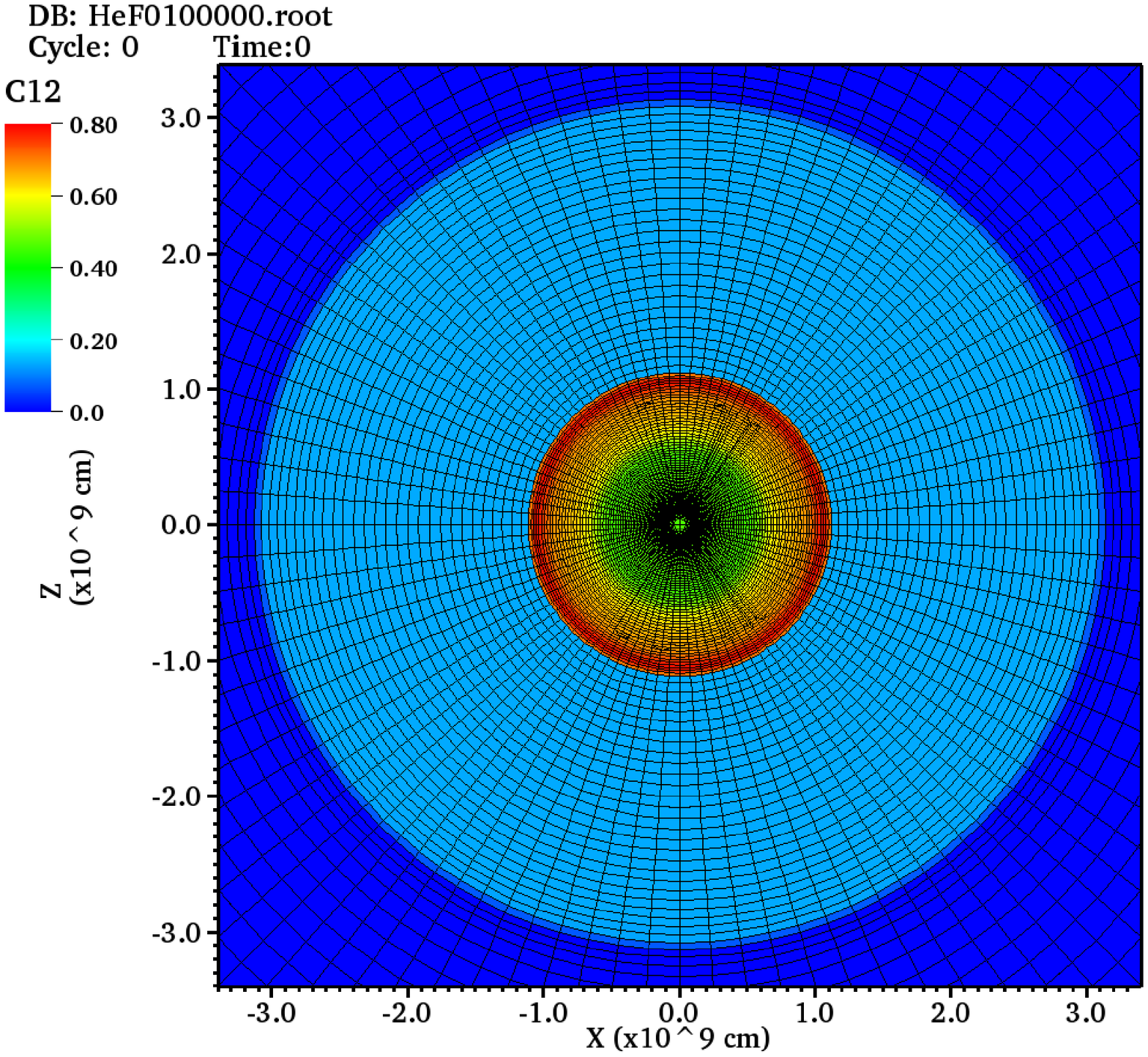} \\
\includegraphics[width=0.55\columnwidth]{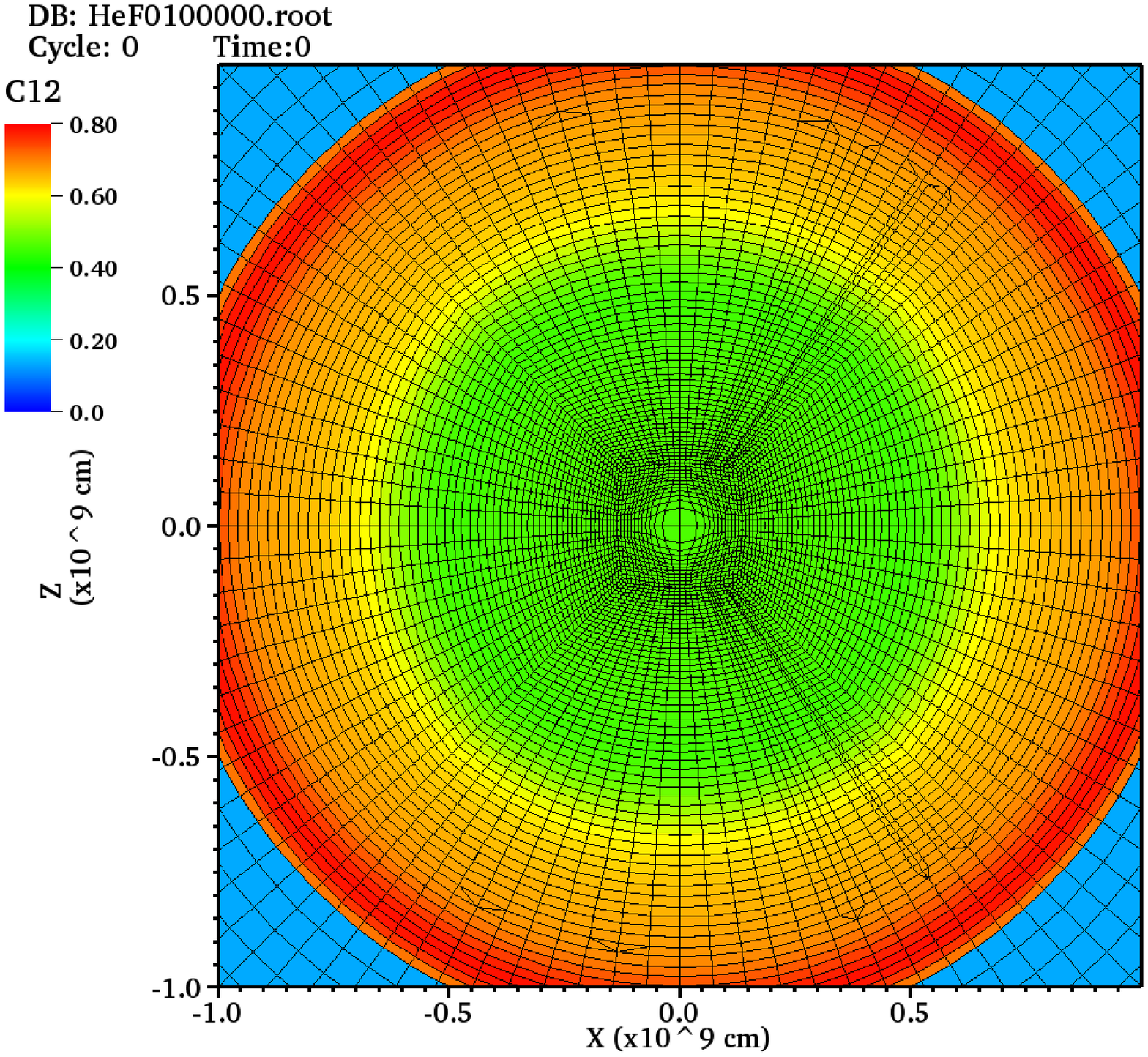}
\caption{The initial mesh of the low resolution run shown in the XZ plane. {\bf Upper panel:} The core and intershell region of the input model, with the mesh structure overlaid. The colour map shows the \el{12}{C} abundance by mass fraction. Above and within the hydrogen burning shell there is essentially no carbon (deep blue). The intershell region has a carbon abundance of around 0.2 (cyan) and the CO core shows variation from red (the approximate location of the He-burning shell) to yellow. {\bf Lower panel:} Close-up of the core region, showing the central cube and the transition from planar to spherical geometry in the arm segments. Note that the asymmetries present on the right hand side of the image are an artefact of the visualisation routine, not a defect in the mesh.}
\label{fig:initialmesh}
\end{figure*}

This 1D model is then mapped into the 3D grid of {\sc djehuty}. We model the interior  of the star from the centre out to just below the convective envelope, at a radius of $4\times10^{11}$\,cm. We have run two separate simulations from the same initial 1D model. In our low resolution run, we set the central cube of the grid to be 20 zones wide. In the arms of the grid, we transistion from cubic to spherical symmetry over 50 zones, and there are 120 radial zones in each arm segment. This gives a total of about 300,000 zones. In the high resolution run, we set the central cube to be 40 zones wide, the transition region to be 80 zones wide and the spherical region contains 200 radial zones. This gives a total of about $2\times10^6$ zones. In both simulations, we have verified that the transition of the mesh to spherical symmetry is complete before the helium burning shell is reached: the whole of our region of interest is simulated in a spherically symmetric grid. A slice through the input model showing the structure of the mesh in the low resolution run is shown in Fig.~\ref{fig:initialmesh}. 

\begin{table*}
\begin{tabular}{lccccc}
\hline
Model & \multicolumn{3}{c}{Width of region in zones} & Total no. & Length of \\
& central cube & transition region & radial arm & of zones & run (hrs) \\
\hline
Low resolution & 20 & 50 & 120 & $2.96\times10^5$ & $4.039$ \\
High resolution & 40 & 80 & 200 & $1.98\times10^6$ & $4.664$ \\
\hline
\end{tabular}
\caption{Details of the two simulations.}
\label{tab:modeldetails}
\end{table*}

We note that the 1D code does not possess the same nuclear network as used in the 3D simulations. In the 1D code, only the energetically important species are included, namely \el{1}{H}, \el{3}{He}, \el{4}{He}, \el{12}{C}, \el{14}{N}, \el{16}{O} and \el{20}{Ne}. The abundances of theses species are used by {\sc djehuty} and the abundances of all the other species in the network are set to zero.

\section{Results}\label{sec:results}

\subsection{The low resolution run}

\begin{figure}
\includegraphics[width=\columnwidth]{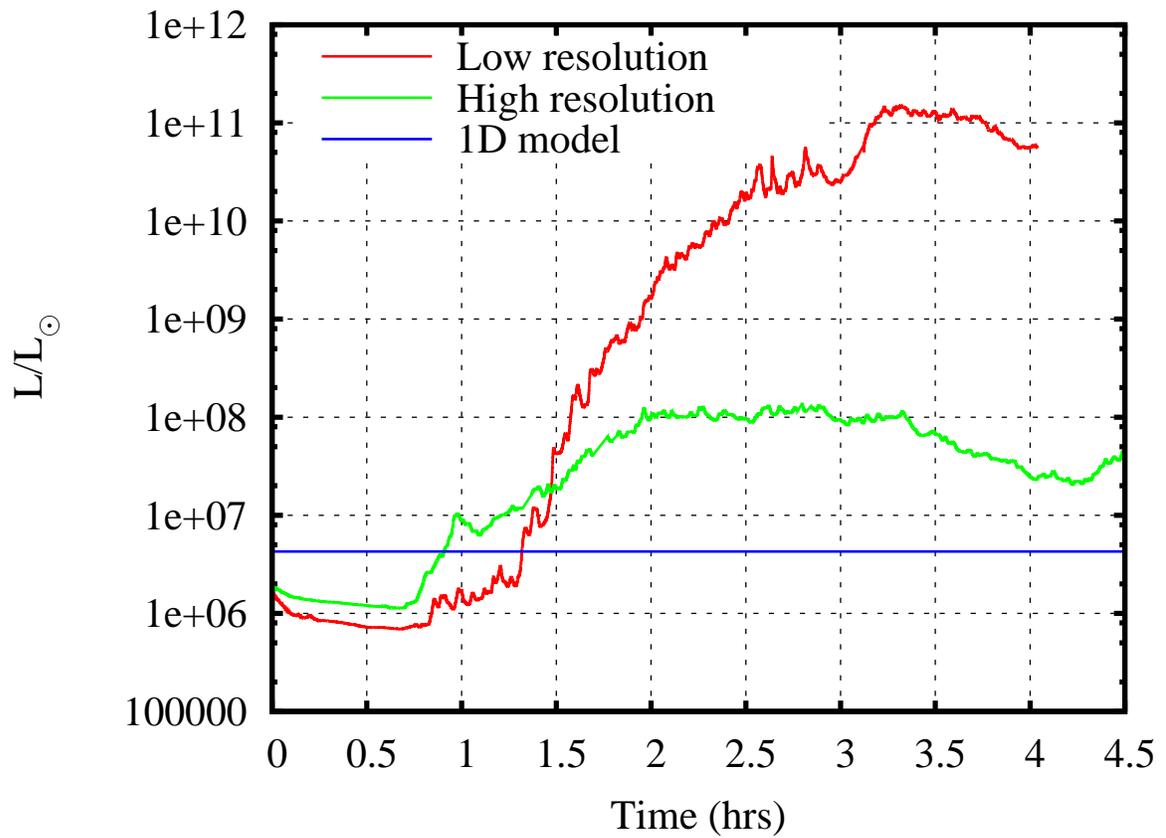}
\caption{Comparison of the total nuclear-burning luminosity between the two runs, together with the luminosity of the 1D input model.}
\label{fig:lumcomparison}
\end{figure}

We have simulated around 4 hours of star time with our low resolution run. This simulation was done using 31 processors and a total of 6.8 CPU years of run time. The total energy generation rate as a function of time is shown in Fig.~\ref{fig:lumcomparison}. Because the input 1D model has no information regarding the velocity field for convective motions, we must wait for the convection to fully develop and the model to settle down. This processes takes around one convective turnover time, which is just under one hour of stellar time in the simulation. During this time, the luminosity is dominated by helium burning, as convective motions have not reached out to the areas of the star that contain hydrogen. At this time the energy generated within the star is about $10^6$\ls, which is in reasonable agreement with the value from the 1D code. At about 1 hour, the convective motions reach the hydrogen-rich regions below the H-burning shell, and the energy output begins to rise. This builds steadily, reaching a peak luminosity of around $10^{11}$\ls. This increase in luminosity comes from the occurrence of proton capture reactions. At the start of the simulation, the total mass of hydrogen in the model\footnote{This is not the same as the total mass of H in the star: recall that our 3D model does not include any of the convective envelope.} is 0.01423\ms, and by the end of the run this has fallen to 0.01405\ms. At the same time, the mass of \el{12}{C} has fallen from 0.28271\ms\ to 0.28175\ms\ (most of the \el{12}{C} is locked in the degenerate core and does not take part in any nuclear reactions). 

\begin{figure*}
\includegraphics[width=0.5\columnwidth]{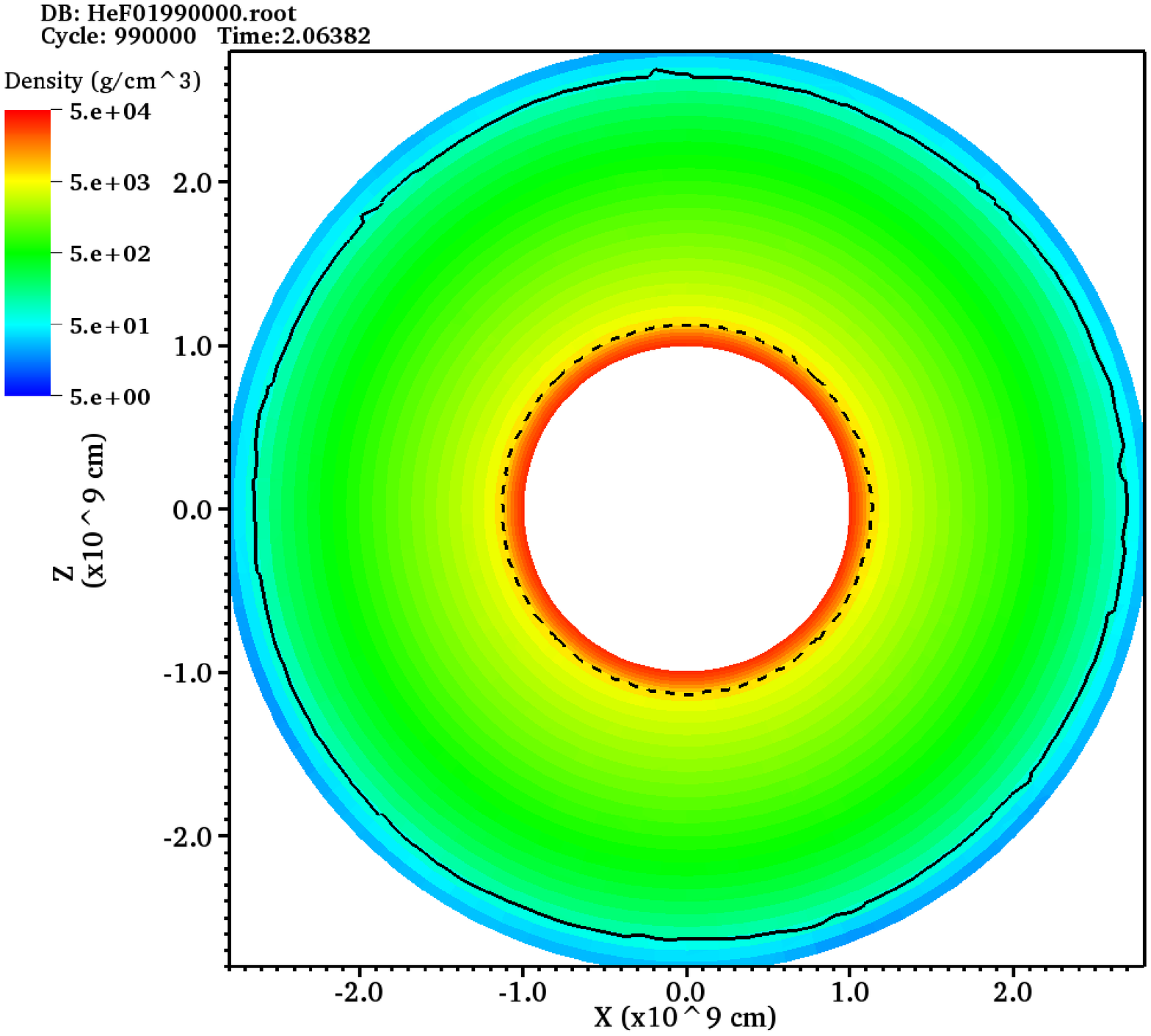}
\includegraphics[width=0.5\columnwidth]{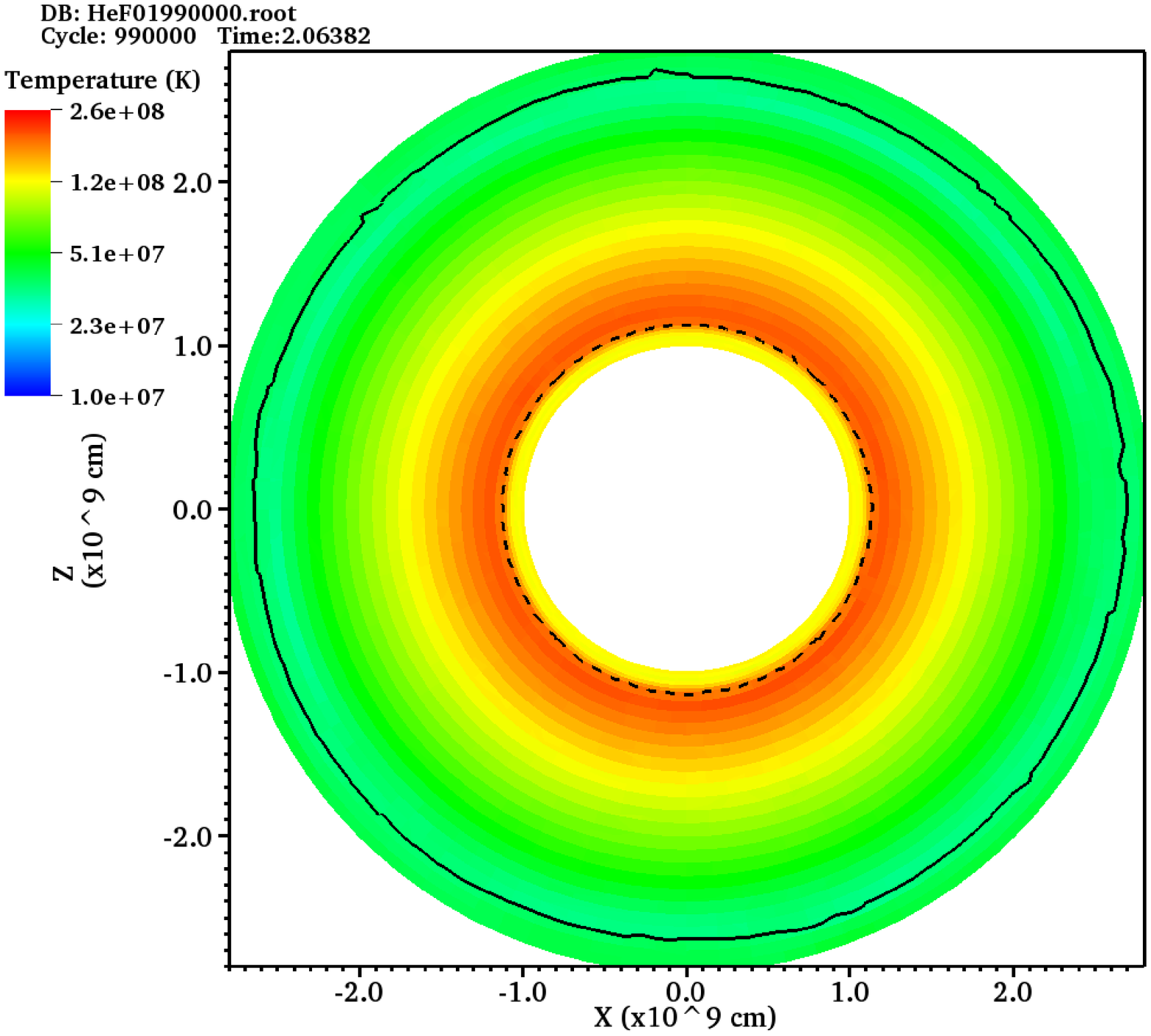}
\includegraphics[width=0.5\columnwidth]{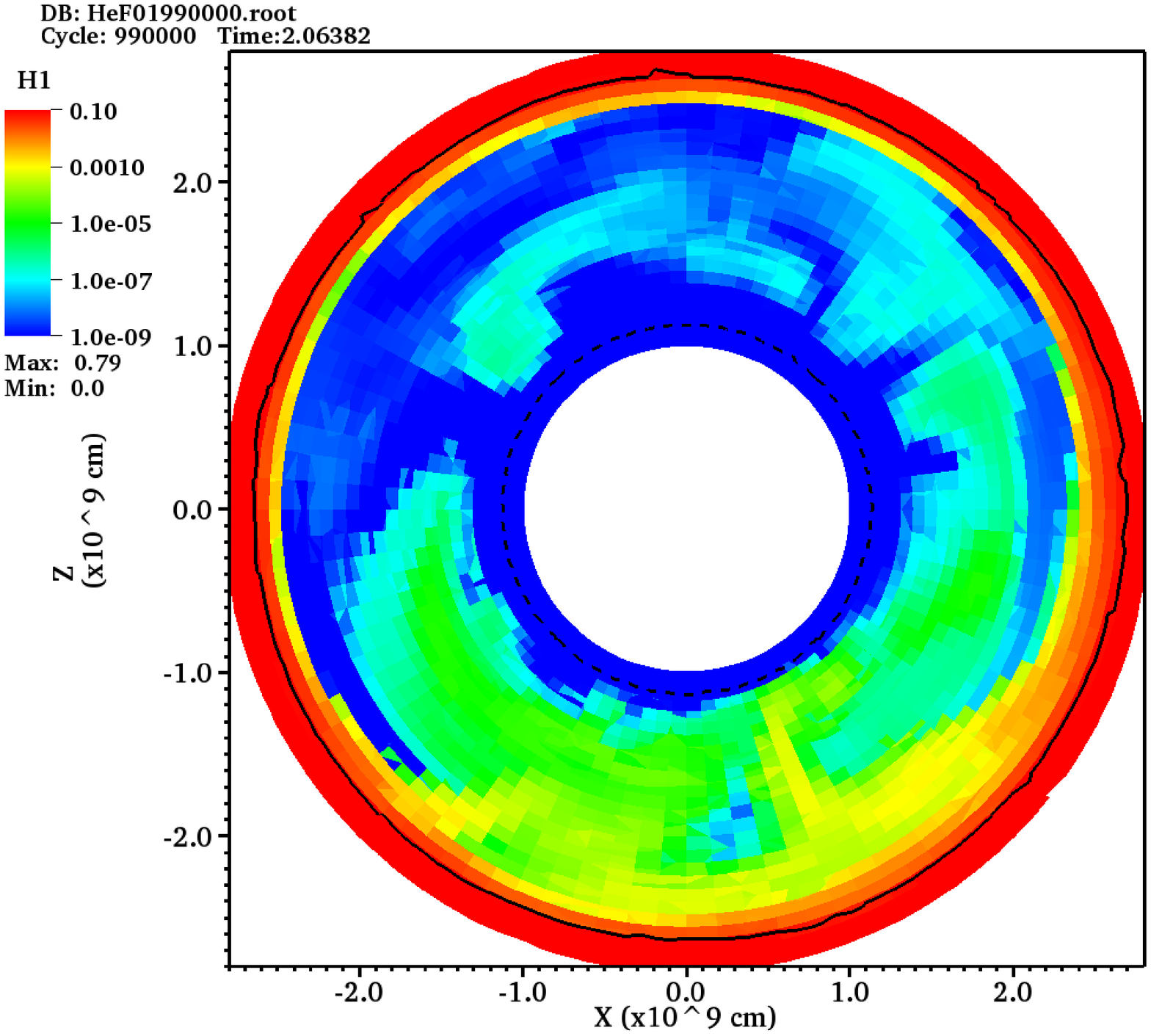}
\includegraphics[width=0.5\columnwidth]{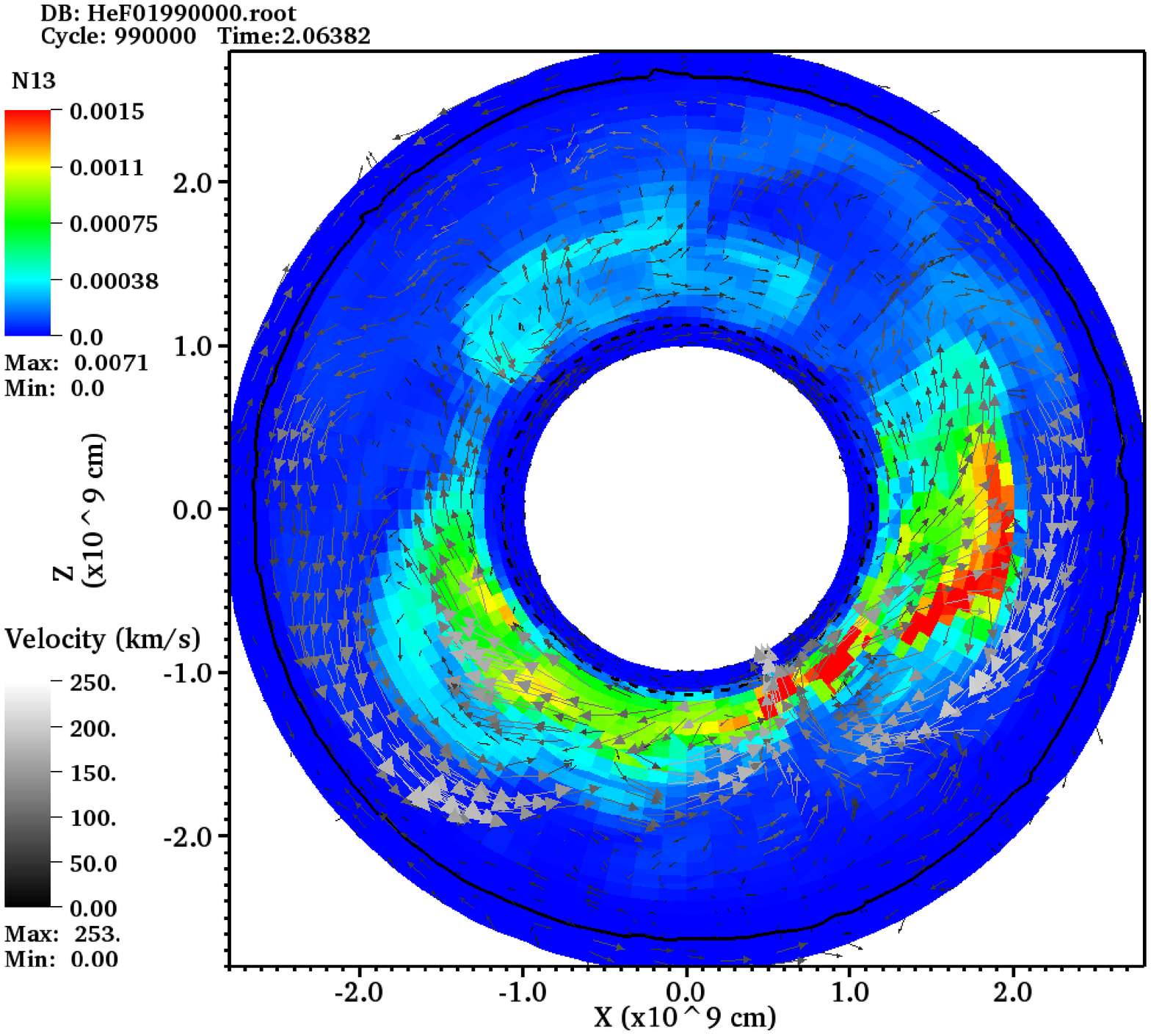}
\caption{Slices along the XZ plane of the low resolution simulation at a time of about 2 hours. In all panels the hydrogen burning shell is denoted by a solid black line and the helium burning shell is denoted by a dashed black line. White regions are simply parts of the simulation that have not been included in the plot for the sake of clarity. {\bf Top left panel:} Colour map showing the density. {\bf Top right panel:} Colour map showing the temperature. {\bf Bottom left panel:} Colour map showing the abundance by mass fraction of hydrogen. {\bf Bottom right panel:} Colour map showing the abundance by mass fraction of \el{13}{N}, with the arrows denoting the velocity of the fluid flow.}
\label{fig:N13low}
\end{figure*}

The upper panels of Fig.~\ref{fig:N13low} show the density and temperature structure of a 2D slice through the XZ plane of the simulation at a time of around 2 hours. Note that they are both very smooth and spherically symmetric. It is the convective motions and their transport of hydrogen that are key to the simulations. Once convective motions have spread throughout the intershell, we find that protons are typically drawn down in narrow, fast flowing plumes. An example is shown in Fig.~\ref{fig:N13low}, which is taken from about 2 hours into the simulation. In the lower left-hand panel, we show the hydrogen abundance in the intershell region. In the lower right-hand panel, we show both the abundance of \el{13}{N} and the velocity field of the model.  \el{13}{N} is a good indicator for where nuclear burning takes place as it is the direct product of the reaction \el{12}{C}\pg\el{13}{N}. Its short half-life (of the order of 10 minutes) means we only see it in active burning regions and as it is mixed away from those regions, prior to it decaying to \el{13}{C}. In the lower left hand panel of Fig.~\ref{fig:N13low}, we see a column of hydrogen with an abundance of around $10^{-4}$ in the lower-right quadrant. A strong downward flow coincides with this hydrogen-rich column as can be seen in the lower right-hand panel. Note that there is little evidence for burning in this downflow: the \el{13}{N} abundance peaks to a very high value only at the very base of this flow and is very low throughout the rest.

This pattern is fairly typical of the proton ingestion we see in the simulation. Fast downflows strip hydrogen from the tail of the hydrogen burning shell and carry it down to very close proximity to the He-burning shell. It is only then that the protons react vigorously, forming a pocket or bubble of \el{13}{N} that is propelled rapidly away from the He burning region. As the newly synthesised \el{13}{N} moves away, it mixes in with the rest of the intershell material. Note that the protons are not burning in flight: they only burn once they have been transported to the helium burning shell. This is contrary to what occurs in the 1D evolutionary models where protons burn as they are transported through the intershell. We shall defer a discussion of the reasons for this to the end of the next section.

\subsection{The high resolution run}

The low resolution run has around $3\times10^5$ zones and was run on 31 processors. Across the intershell we have around 20 zones in the radial direction and around 80 zones in the azimuthal directions (recall the central cube had 20 zones per side and this sets the angular resolution). We ran a second simulation with closer to $2\times10^6$ zones. This improves the radial and angular resolution by a factor of about two, and we now have about 40 zones radially across the intershell. The increase in the number of zones comes at the cost of having to increase the number of processors (144 were used here) and the increased resolution reduces the (Courant-limited) timestep.

Two modifications to the code were made prior to the high resolution run. First, the code was made to output the energy generation from various burning reaction groups (e.g. the pp-chains, the CNO cycle, the triple-$\alpha$ reaction, etc.). Secondly, the nuclear reaction network was updated to include neutron producing reactions and neutrons were added as a species. Specifically, we added the reactions \el{13}{C}\an\el{16}{O}, \el{17}{O}\an\el{20}{Ne} and \el{22}{Ne}\an\el{25}{Mg}. The rates used were taken from \citet{1988ADNDT..40..283C}. At present, neutrons do not participate in any further reactions (i.e. they are not absorbed by any of the species). We plan to update this in future work when we study the nucleosynthesis in detail.

\begin{figure}
\includegraphics[width=\columnwidth]{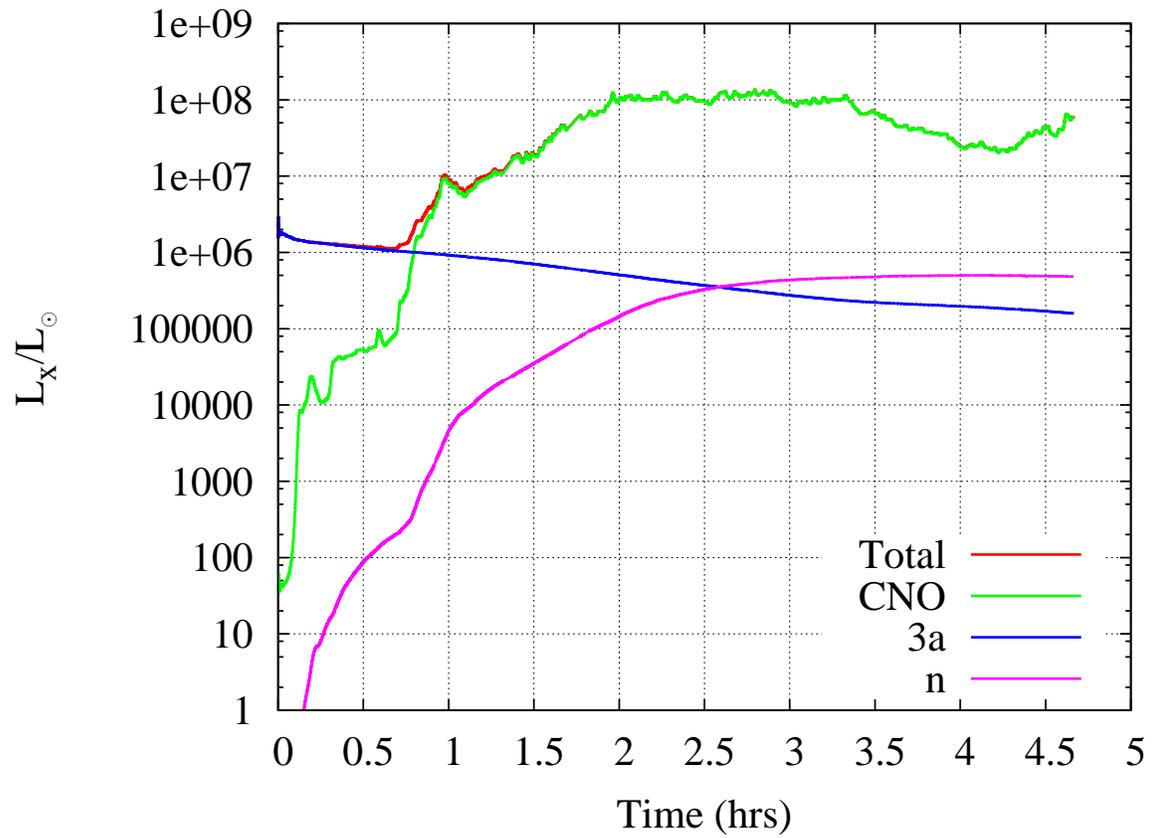}
\caption{Luminosity as a function of time for various burning sources in the high resolution model.}
\label{fig:luminosities}
\end{figure}

Fig.~\ref{fig:luminosities} shows the luminosity from CNO reactions, triple-$\alpha$ burning and neutron-generating reactions (the dominant neutron source is the \el{13}{C}\an\el{16}{O} reaction). After roughly one convective turn-over time, the CNO burning luminosity becomes the dominant source of energy (as is found in 1D simulations which have substantial ingestion of protons, such as those of \citealt{2009PASA...26..139C} for thermal pulses and \citealt{2010A&A...522L...6C} for the core helium flash). After about 2.5 hours, energy generation by neutron producing reactions exceeds energy production via the triple-$\alpha$ reaction. However, the neutron-producing reactions remain of little energetic importance, with energy generation from CNO burning being around 2 orders of magnitude greater.

A comparison between the total luminosity from both the low and high resolution runs is shown in Fig.~\ref{fig:lumcomparison}. The high resolution model has a greater luminosity than the low resolution model in the early, settling down phase (up to about 0.7 hours) by about 50\%. In the high resolution run, the total luminosity increases dramatically from 0.7 hours, reaching $10^7$\ls\ by around 1 hour. In contrast, the low resolution simulation shows some low-level fluctuations in the luminosity after the settling down period and a sharp increase in luminosity does not occur until after 1.25 hours.

The sudden increase in the total luminosity in the high resolution model is a result of the ingestion of protons into the intershell convection zone. This can be seen when one looks at the contributions to the total luminosity, as shown in Fig.~\ref{fig:luminosities}. CNO cycle reactions become the dominant energy source just prior to 0.8 hours -- a clear signature that proton ingestion has begun. Before this point, the triple-alpha reaction was the dominant energy source. Note that we have substantial activation of the \el{13}{C}\an\el{16}{O} reaction in the model as the luminosity from neutron-producing reactions has reached over $10^4$\ls. The \el{13}{C} is generated when \el{12}{C} captures a proton to form the unstable nucleus \el{13}{N} which then $\beta$-decays to form \el{13}{C}.

\begin{figure*}
\includegraphics[width=0.6\columnwidth]{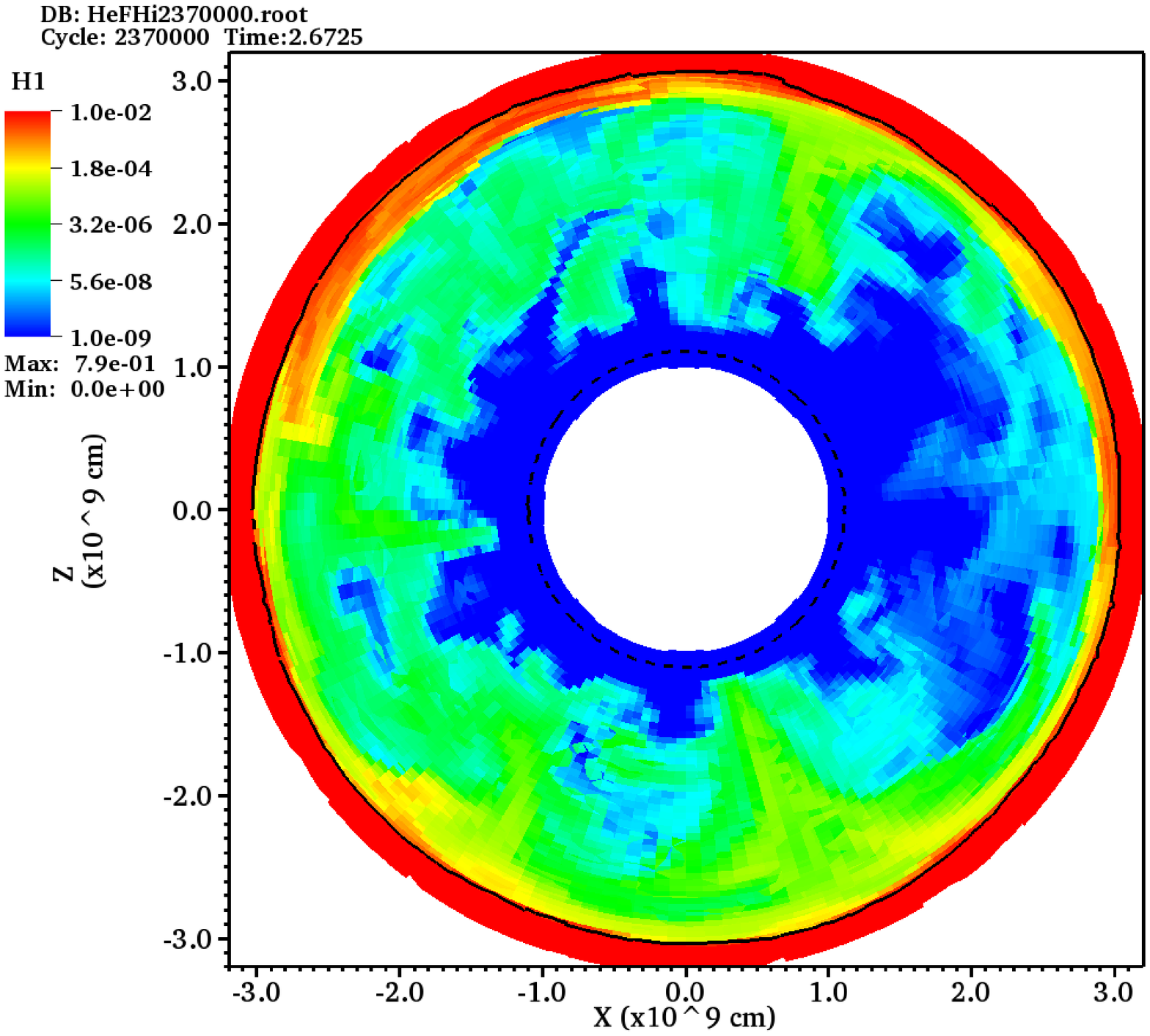} \\
\includegraphics[width=0.6\columnwidth]{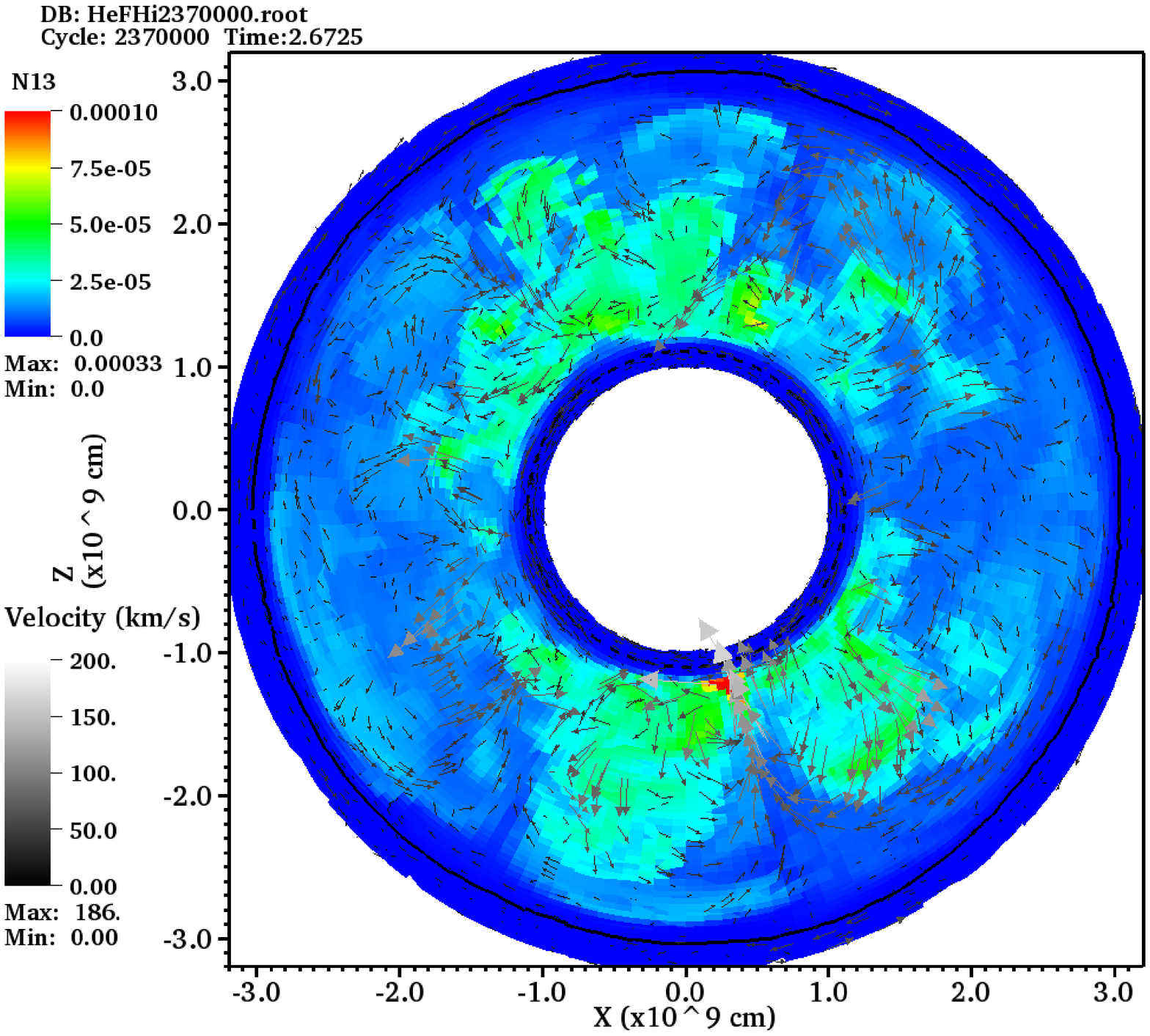}
\caption{Slices along the XZ plane of the high resolution simulation at a time of about 2.7 hours. In both panels the hydrogen burning shell is denoted by a solid black line and the helium burning shell is denoted by a dashed black line. White regions are simply parts of the simulation that have not been included in the plot for the sake of clarity. {\bf Upper panel:} Colour map showing the abundance by mass fraction of hydrogen. {\bf Lower panel:} Colour map showing the abundance by mass fraction of \el{13}{N}, with the arrows denoting the direction and magnitude of the fluid flow. Animated versions of these plots (showing the evolution over the time range 1.41-4.22 hours) are available from the journal's website.}
\label{fig:N13high}
\end{figure*}

With regards to the turbulent convection, we see similar structures in the low and high resolution runs. Fig.~\ref{fig:N13high} shows slices through the XZ plane of the high resolution run. We again see plume-like ingestion of hydrogen (see the upper panel of Fig.~\ref{fig:N13high}). However, in the high resolution runs we see more individual plumes, possibly as a result of the convective motions being better resolved. The high resolution simulation shows greater structure in its velocity field, with more individual convective cells being present. The high resolution simulation also has lower convective velocities.

\begin{figure*}
\includegraphics[width=0.68\columnwidth]{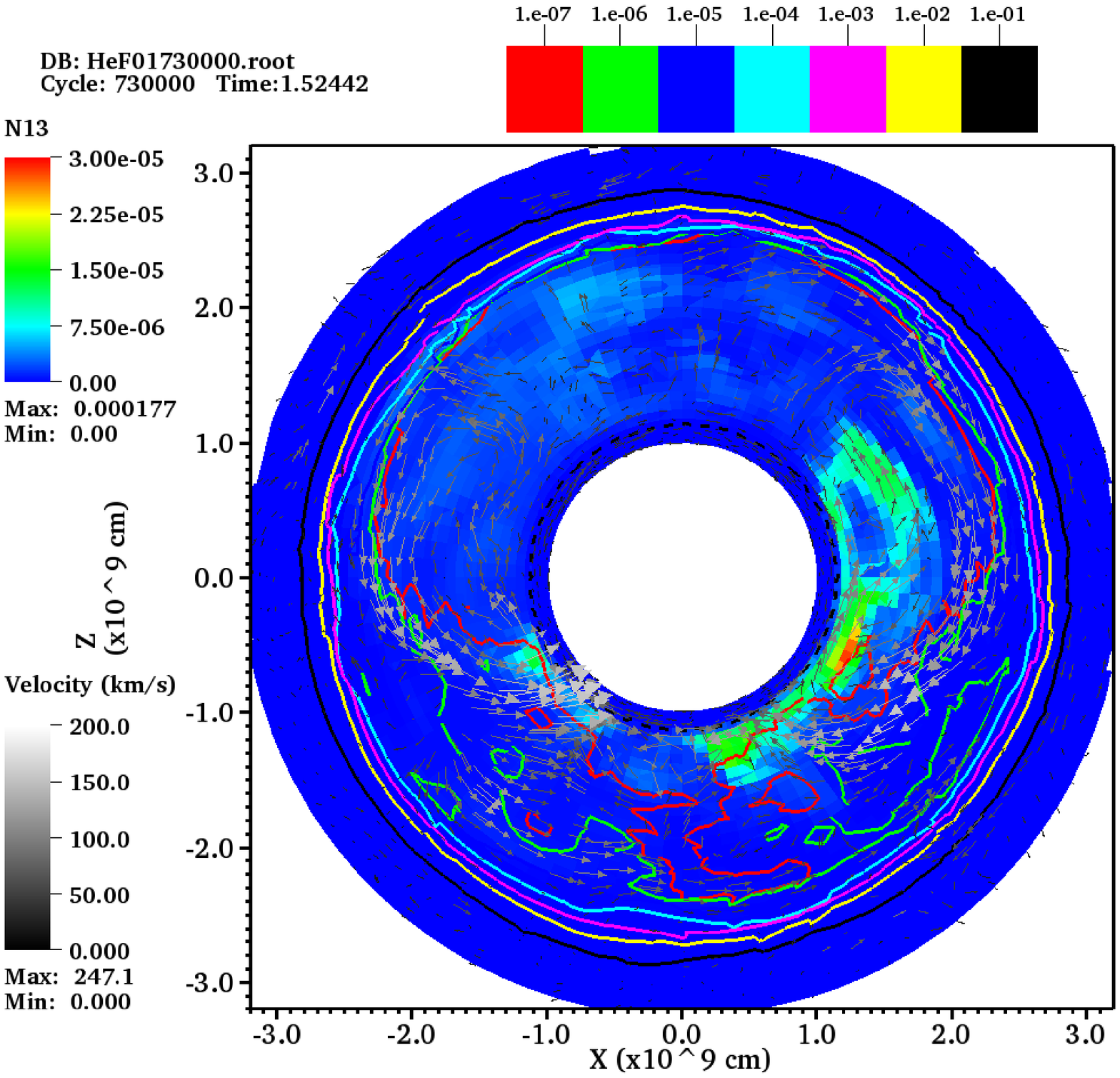} \\ 
\includegraphics[width=0.68\columnwidth]{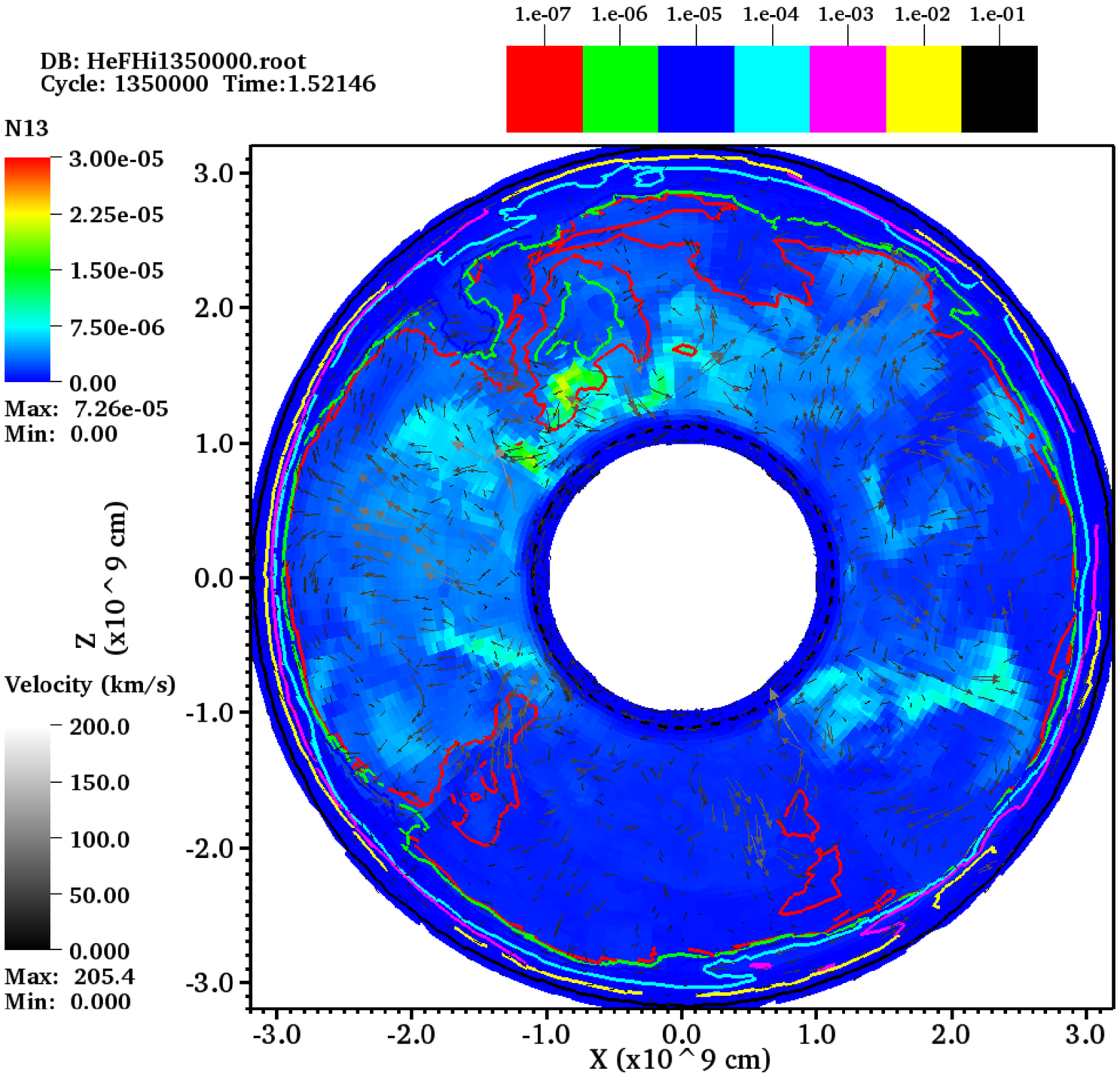}
\caption{2D slice through the model for the low resolution (upper panel) and high resolution (lower panel) runs, showing the \el{13}{N} abundance (colour map), contours of \el{1}{H} and velocity (grey arrows). The plots are as co-incident in time as the output allows (the difference is about $10$\,s).}
\label{fig:N13H1comparison}
\end{figure*}

Fig.~\ref{fig:N13H1comparison} shows slices through both the low and high resolution runs at a time of 1.5 hours. The first thing to note is that the low resolution run is slightly more compact than the high resolution run. The $X_\mathrm{H}=0.1$ contour falls just within $3\times10^9$cm from the centre, whereas in the high resolution run the same contour lies just outside this radius (compare the upper and lower panels of Fig.~\ref{fig:N13H1comparison}). This contour also seems to be more spatially separated from the $X_\mathrm{H}=0.01$ contour in the low resolution model. This is because the low resolution run mixes down more material from the hydrogen-rich regions, on account of it having a thicker boundary layer and more entrainment because of its lower resolution (see the discussion at the end of this section). Contours of a given hydrogen abundance therefore appear at smaller radii.

The peak \el{13}{N} abundances seem comparable in both runs at this point. However, the low resolution run has more localised pockets of high \el{13}{N} abundance while in the high resolution simulation, the \el{13}{N} seems more widely spread. The low-resolution simulation seems to have a much greater degree of proton ingestion at this point. The $X_\mathrm{H}=10^{-7}$ contour (the red line in Fig.~\ref{fig:N13H1comparison}) has reached almost to the He burning shell in the lower half of this simulation and there is also significant distortion of the $X_\mathrm{H}=10^{-6}$ contour (denoted in green). In the high resolution simulation there is less overall distortion of the hydrogen contours, though there are significant downward plumes in evidence (notably on the top- and bottom-left of the plot). In addition, there seem to be differences in the flows in the two simulations. The low resolution simulation displays significantly faster flows (note the large, whitish arrows in the lower half of the plot). There seems to be significant flow around most of the intershell, with two large eddies accounting for the flow in the lower half of the image. In contrast, the flow in the high resolution simulation seems to be much slower and split into more, smaller cells.

\begin{figure*}
\includegraphics[width=0.5\columnwidth]{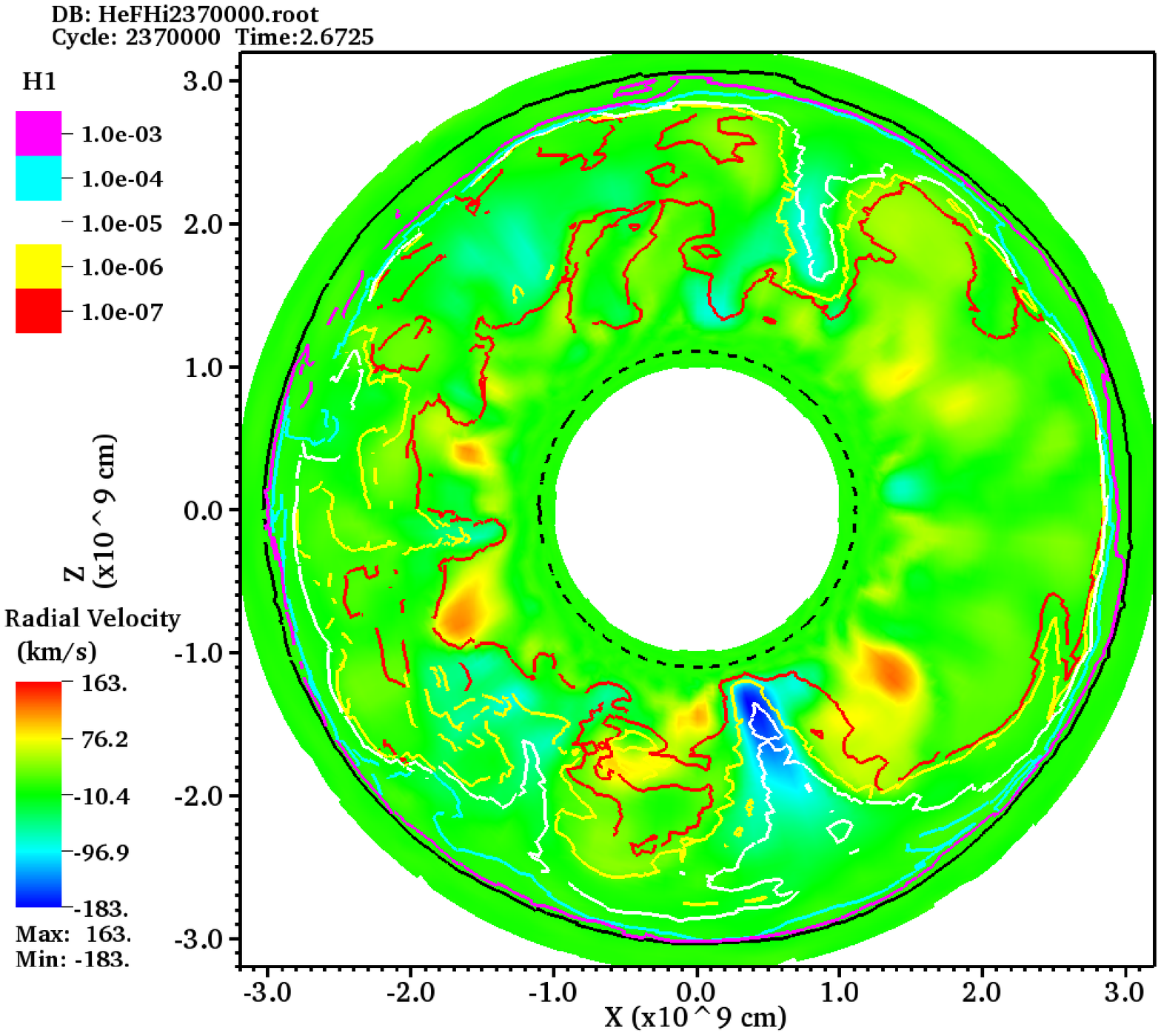}
\includegraphics[width=0.5\columnwidth]{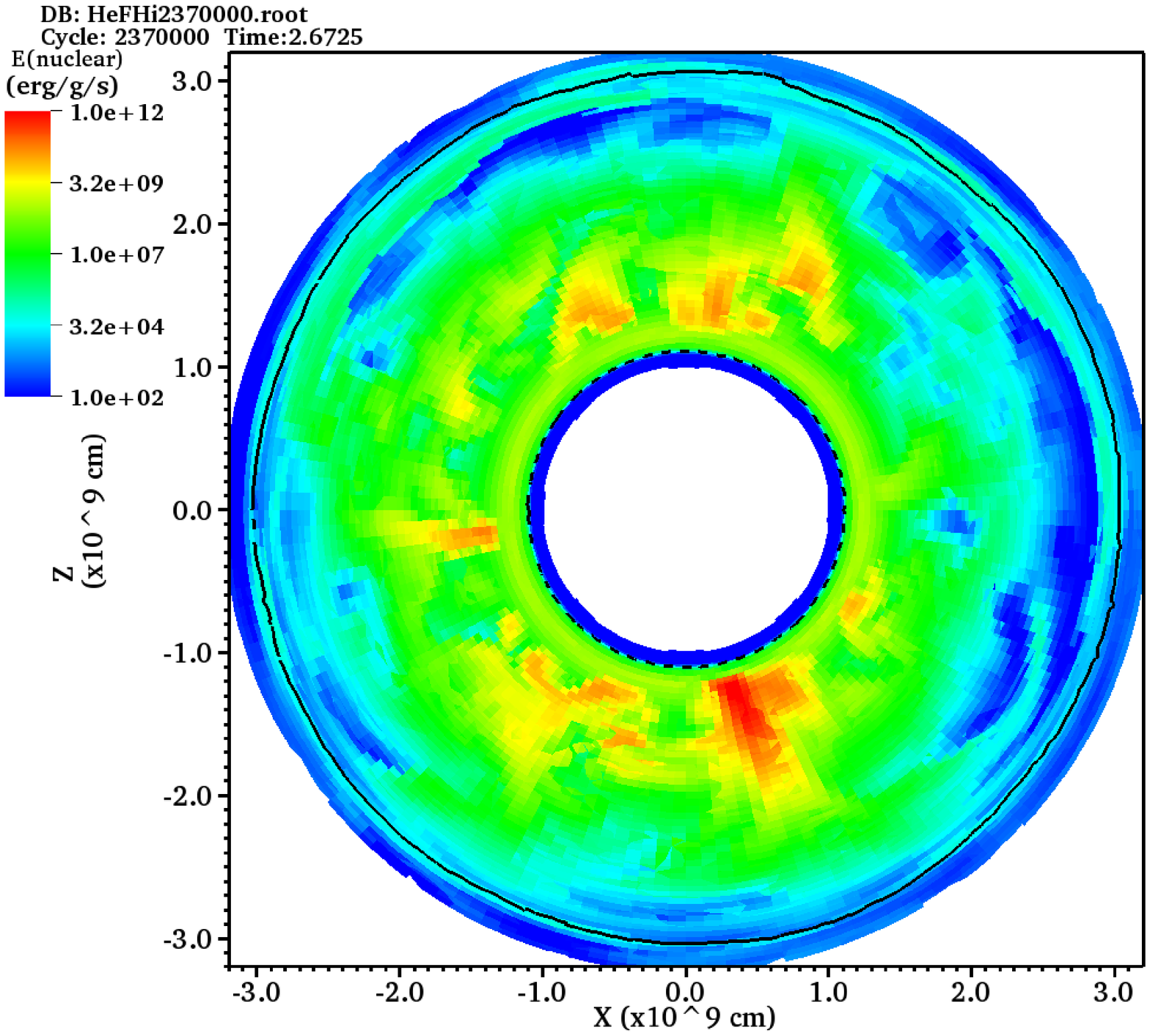}
\includegraphics[width=0.5\columnwidth]{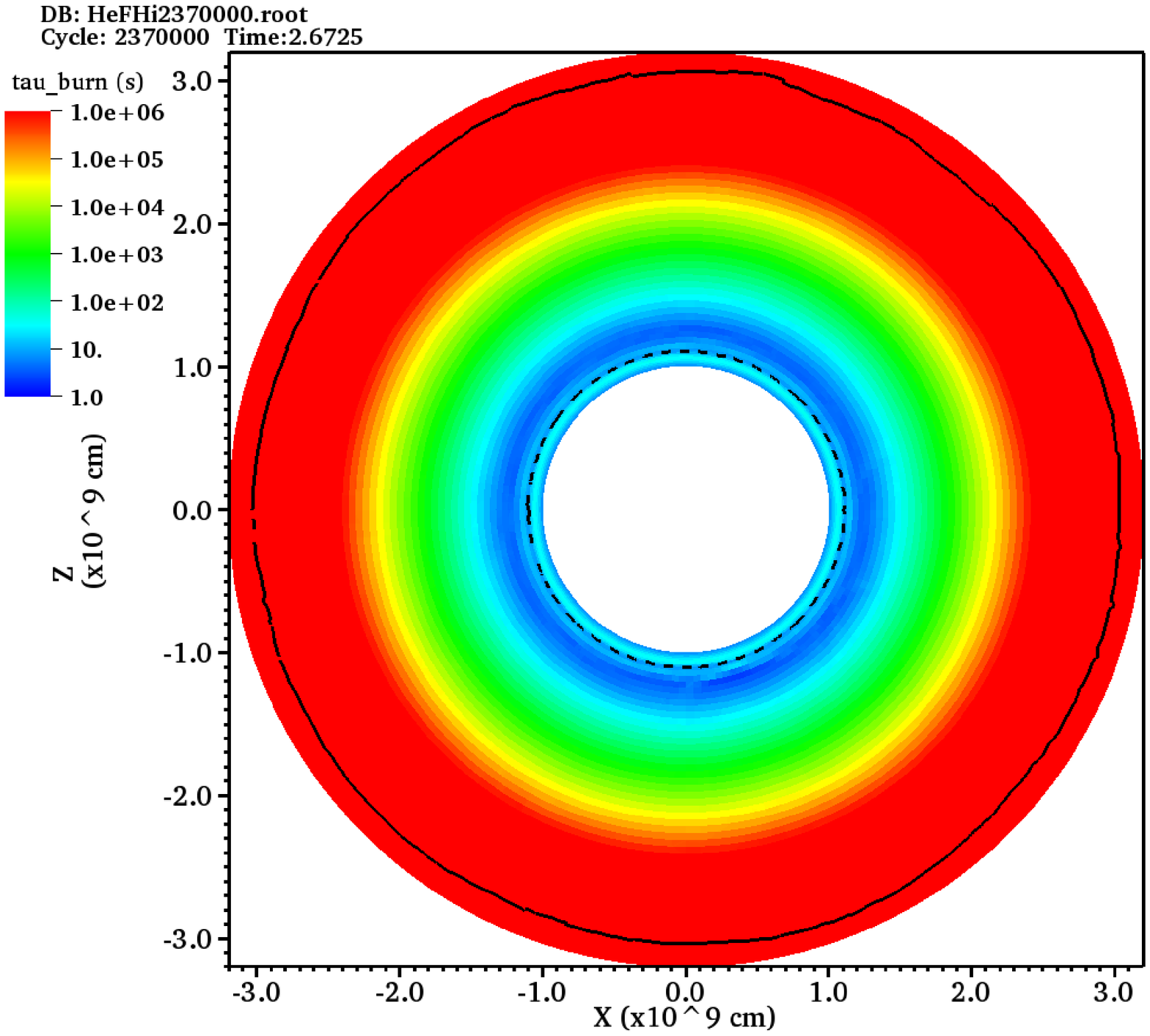}
\includegraphics[width=0.5\columnwidth]{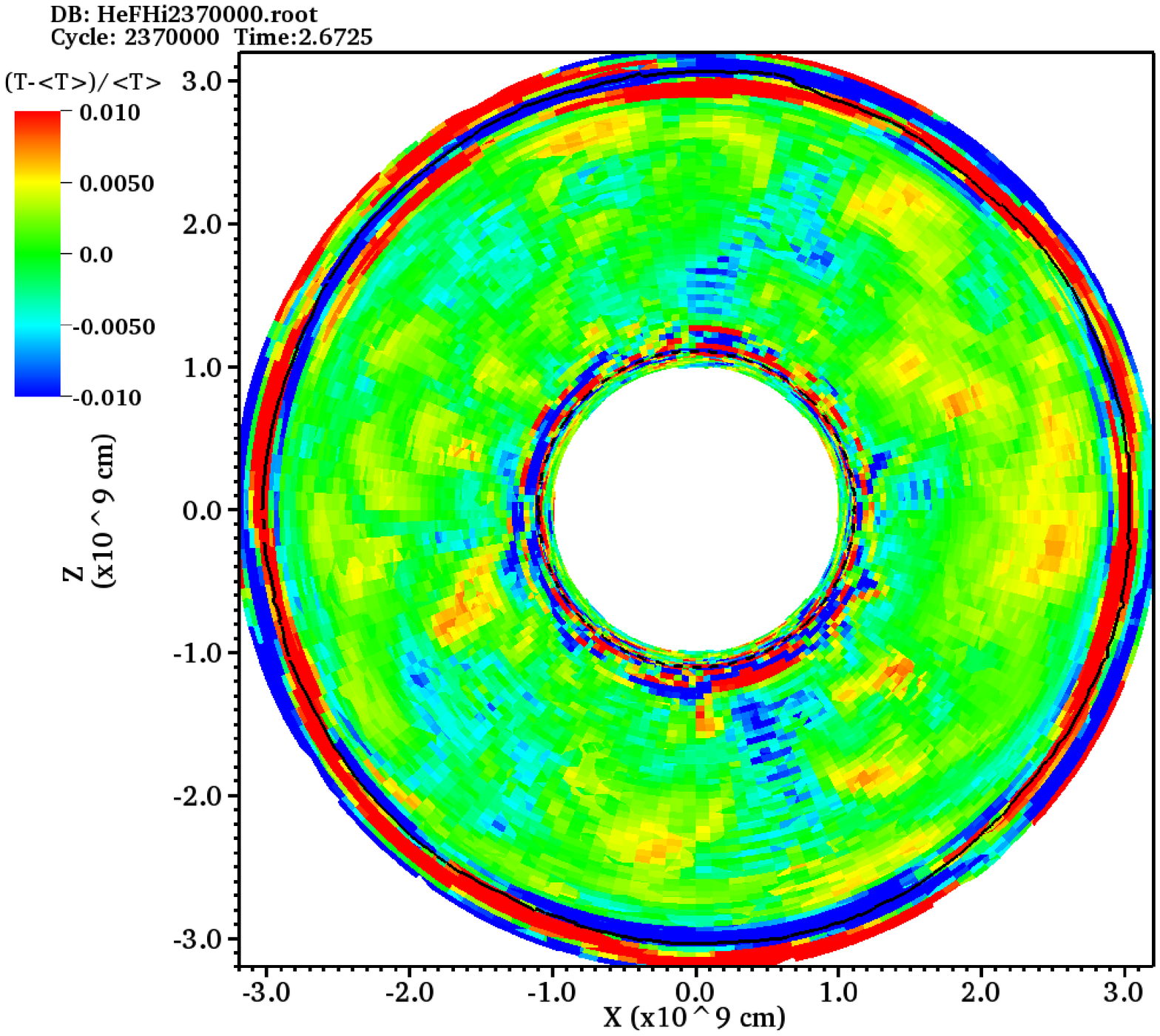}
\caption{Slices along the XZ plane of the high resolution simulation for the same timestep as in Fig.~\ref{fig:N13high}.  {\bf Top left panel:} Colour map showing the radial velocity of the matter in the simulation, along with contours of the hydrogen abundance. {\bf Top right panel:} Colour map showing the energy generation in the model. {\bf Bottom left panel:} Colour map showing the timescale for the reaction \el{12}{C}\pg\el{13}{N}. Note this has been capped at a maximum of $10^6$\,s for clarity. {\bf Bottom right panel:} Colour map showing the temperature contrast (i.e. temperature deviation from the average on a spherical shell) for the model. Animated versions of the two plots in the upper panels (showing the evolution over the time range 1.41-4.22 hours) are available from the journal's website}
\label{fig:plumeplots}
\end{figure*}

The upper left panel of Fig~\ref{fig:plumeplots} shows the radial velocity for a slice through our simulation. In the lower right quadrant, there is a strong downward plume, with material that is rich in hydrogen flowing at speeds of over 100\,km/s. This rapid flow means that the protons do not have time to burn in flight, and in fact strong nuclear burning only occurs once the protons have been transported to close to the helium burning shell (see the upper right panel of Fig~\ref{fig:plumeplots}). Following \citet{2011ApJ...727...89H}, we may define a burning timescale for the reaction \el{12}{C}\pg\el{13}{N} as:
\be
\tau_\mathrm{burn} = {12 \over X_{^{12}\mathrm{C}} \rho \mathrm{N_A} \langle\sigma v\rangle}
\label{eq:timescale}
\ee
where $X_{^{12}\mathrm{C}}$ is the abundance by mass fraction of \el{12}{C}, $\rho$ is the density, $\langle\sigma v\rangle$ is the cross-section for the reaction and $N_A$ is Avagadro's number. This burning timescale is plotted in the lower panel of Fig.~\ref{fig:plumeplots}. Burning will only take place when the timescale for the transport of protons becomes comparable to the burning timescale. With a downflow velocity of around 100\,km/s, the entire width of the intershell (around $2\times10^4$\,km) would be traversed in just 200\,s. Burning timescales this short are only reached close to the helium burning shell, and so we only obtain strong hydrogen burning in such regions as can be seen in the upper-right panel of Fig.~\ref{fig:plumeplots}. Perhaps surprisingly, this strong burning does not have a substantial effect on the temperature structure of the model. In the lower-right panel of Fig.~\ref{fig:plumeplots}, we plot the temperature contrast of the model. This is the deviation in the local temperature, $T$,  from the average temperature over a spherical shell with the same radius, $\langle T\rangle$, measured relative to that average, i.e. $(T-\langle T\rangle)/\langle T\rangle$. Within the intershell, the temperature deviates by no more than around 1 per cent from the spherical average for that radius.

One may question whether we have enough resolution in the high resolution run. The two runs seem quite dissimilar in behaviour. What would happen if we increased the resolution further? We are working on a higher resolution run at present, but it is unlikely to provide a large enough change in resolution. In the 1D evolution code, we use around 600 mesh points to cover the intershell region \citep{2004MNRAS.352..984S}, whereas there are only about 40 cells in the radial direction between the helium and hydrogen burning shells. An order of magnitude increase in this resolution, although desirable, is not possible at present.

The less than optimal resolution of our simulations means we should place two caveats on our results. At the interface between the convective intershell and the hydrogen rich regions, we are likely to be overestimating the degree of entrainment. Low resolution will lead to a broader entrainment layer with greater viscosity and consequently to more entrainment. That this is so can be seen in the difference between our low and high resolution simulations. The low resolution simulation ingests (entrains) more hydrogen, reaching high burning luminosities as a consequence. Until we have properly resolved the boundary layer, we cannot quantify the mass entrainment rate and so we refrain from attempting to do so.
 
In addition, the simulation may not represent truly turbulent convective motions in the intershell region. Because {\sc djehuty} does not include sub-grid scale physics, the viscosity present in the simulations is numerical and based upon the grid size used. If the grid size is not sufficiently small, we will not be able to resolve eddies on the smallest scales and our convective flows will not be fully turbulent. Consequently we may underestimate the degree of mixing that takes place. One must therefore caution that the results of this study may not be borne out by higher resolution studies. It is well known that as one moves from the laminar to the turbulent flow regime the nature of the flow patterns can change dramatically \citep[see e.g.][]{2000ApJ...532..593M}. It is possible that at higher resolution, we would not retain such fast-flowing, coherent, downward plumes because of additional turbulence.

We can make a crude estimate of the Reynolds number (Re) of our simulations. \citet{davidson2004} gives an expression for this quantity in terms of the number of zones (N) in the computational domain and the ratio of the size of the largest eddy (l) to the physical size of the region being simulated (L). His equation (7.3) states:
\be
\mathrm{Re} \approx \left(l\over L\right)^{4\over 3}N^{4\over9}.
\ee
For both our simulations, the size of the largest eddies is comparable to the size of the intershell region so that $l/L$ is of the order of unity. Consequently our estimate of the Reynolds number scales (slowly) with the number of zones in the intershell region. For the low resolution simulation, the intershell has about 20 zones in the radial direction and each of the six arm segments has a cross section of 20$\times$20 zones. We therefore have roughly 48,000 zones in this region which yields Re~$\approx120$. For the high resolution simulation, we have about 45 zones radially across the intershell and each arm has a cross section of 40$\times$40 zones, yielding a total of 432,000 zones. This gives Re~$\approx320$. These numbers suggest that our simulations are turbulent, but only marginally so. To be certain we are in the fully turbulent regime, we would need at least an order of magnitude more zones in our intershell.

\subsection{Comparison to the models of Herwig et al.}

Our simulations display qualitative agreement with those of \citet{2011ApJ...727...89H}, despite their simulations not including burning of the ingested hydrogen. We also find that hydrogen ingestion takes place where upwelling cells meet, and that the ingested material is carried down in narrow, fast flowing plumes. Our radial velocities reach mean values of 30-40\,km/s (whereas mixing length theory predicts velocities of 1-2\,km/s for our 1D input model), compared to 12\,km/s in the Herwig et al. simulations. There are likely two contributions to this difference. First, we have greater energy generation in our model, leading to stronger convective motions. Their model was driven by a total luminosity of $4.2\times10^7$\ls, whereas we have a total luminosity reaching up to as much as $10^8$\ls\ in the high resolution case, owing to the release of energy from hydrogen burning which is not included in the Herwig et al. simulations. Secondly, the Herwig et al. simulations have much higher resolution than ours and are more likely to be truly turbulent.

\section{Comparison to the 1D models}\label{sec:1d}

Perhaps the most striking difference between the 1D and 3D simulations is in the hydrogen burning luminosity ($L_\mathrm{H}$). At the timestep that was used as the input model for the 3D simulations, $L_\mathrm{H}\approx 5.5\times10^2$\ls. The proton ingestion is relatively minor at this point, as can be seen in Fig.~\ref{fig:1Dabunds}. The intershell convective region extends from around $1.2\times10^9$\,cm to around $3\times10^9$\,cm. The H abundance is never greater than $10^{-5}$ in this region and it drops below $10^{-12}$ around $2\times10^{9}$\,cm. There is no significant proton abundance below this. This line marks a fairly good approximation of the lower envelope of the hydrogen abundance seen in the 3D simulation. However, the 3D simulation\footnote{Unless otherwise specified, we are referring to the high resolution simulation.} shows significant variation in the H-abundance in the convective region, reaching as high as $X_\mathrm{H}\approx10^{-4}$ throughout, even down to the He-burning shell. This accounts for the significantly higher H-luminosity found in this simulation.

\begin{figure}
\includegraphics[angle=270,width=\columnwidth]{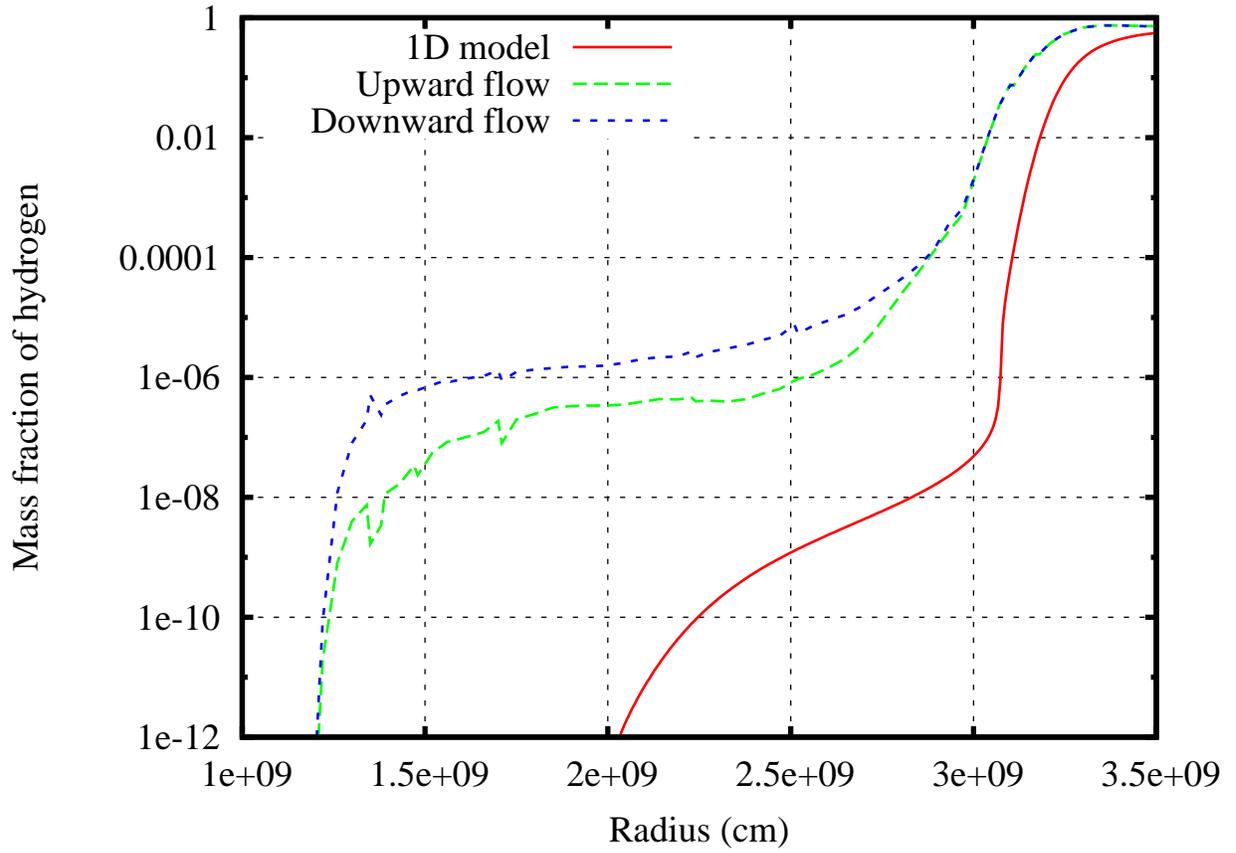}
\caption{Hydrogen abundance by mass fraction for the models. The hydrogen abundances for the up and down streams from the 3D model are mass-weighted averages for cells at a given radius.}
\label{fig:Habundances}
\end{figure}

If we take a mass-weighted average over radial shells in the 3D model, and also separate the cells according to their radial motion (i.e. in to up- and downflows) then we obtain the profiles shown in Fig.~\ref{fig:Habundances}.  The H abundance in the downflow is clearly greater than that in the upward flow, perhaps by as much as one order of magnitude. It is also clear that the average abundance of H in the convection zone of the 3D model is far higher than in the 1D model and this is the reason for the much higher hydrogen burning luminosity. We note that close to the upper edge of the convective zone, the hydrogen profile in the 3D model has a similar shape to the 1D model which uses diffusive mixing, i.e. it looks as though the mixing here is more diffusive in nature. \citet{2011ApJ...727...89H} also found this to be the case in their simulations. However, as we move away from the boundary, the abundance profiles in the 3D model are noticeably flatter, before finally falling off sharply as the burning shell is approached. We believe this profile would be better represented by an advective mixing scheme, rather than the diffusive scheme employed by the 1D evolution code. This conclusion underlines the point made by \citet{2011ApJ...733...78A} that convection is advective in nature, rather than diffusive.

\begin{figure}
\includegraphics[angle=270,width=\columnwidth]{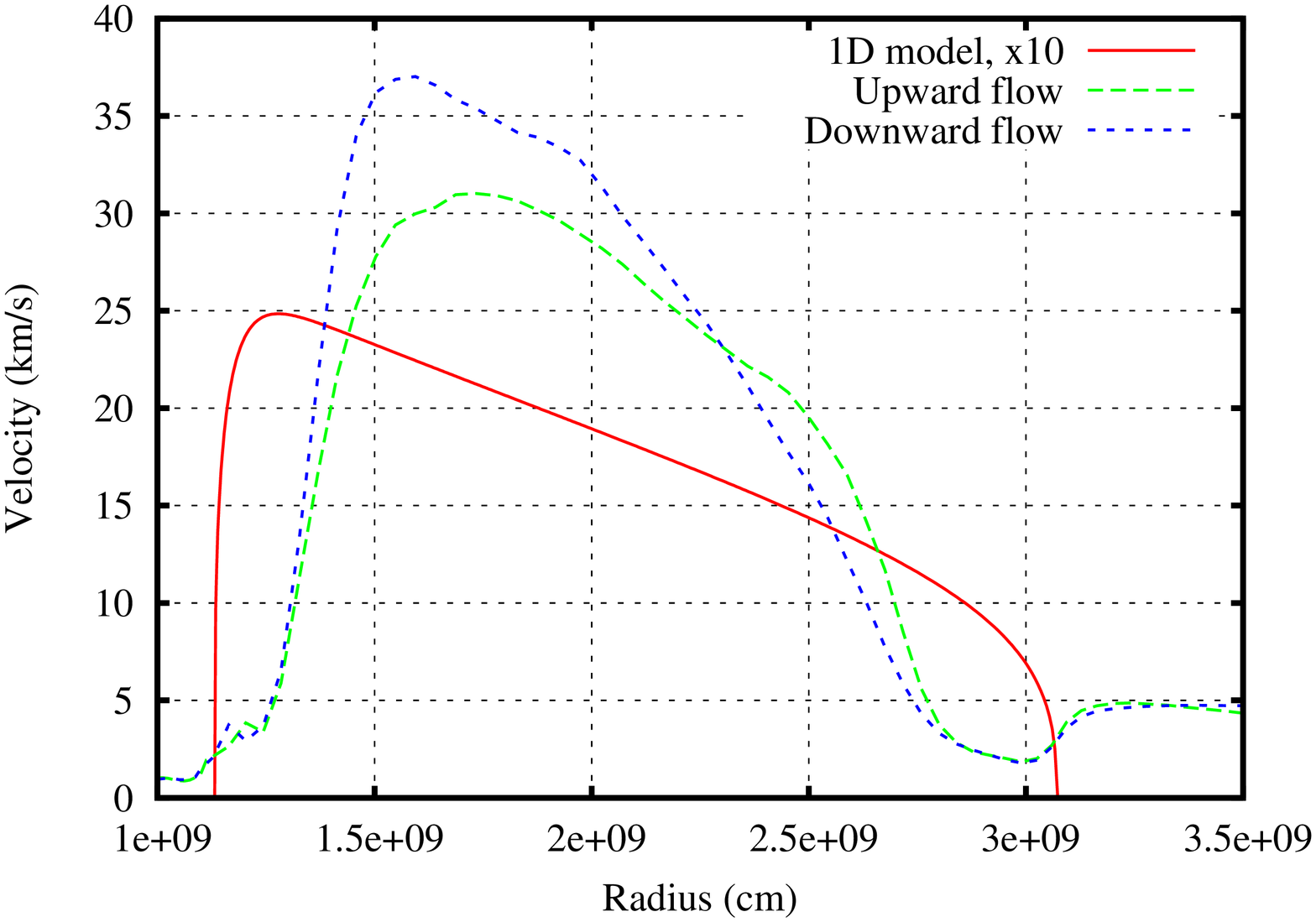}
\caption{Convective velocities as a function of radius for the models. The convective velocity for the 1D model is the MLT value and has been multiplied by a factor of 10 so that its shape can be easily distinguished. The velocities for the up and down streams from the 3D model are a mass-weighted average of the radial velocity for cells at a given radius.}
\label{fig:velocities}
\end{figure}

In Fig.~\ref{fig:velocities} we plot the mixing length velocity for the 1D model, along with the mass-weighted average velocities in the up- and downstreams. We note that the 1D MLT velocity is over an order of magnitude {\it slower} than the convective motions in the 3D simulation, a result in agreement with the simulations of \citet{2011ApJ...727...89H}. We also note that there is substantial asymmetry between the up- and downflows, particularly towards the base of the convective region. Below a radius of $2\times10^9$\,cm, the downflow velocity can exceed the upflow velocity by around 7\,km/s. Above this point, the two flows have comparable velocity and the upflow is faster than the downflow above a radius of $2.2\times10^9$\,cm. Note also that this is for the {\it average} upward and downward flows. The velocity of fluid elements in these flows can be significantly in excess of these velocities. Peak velocities of over 100\,km/s regularly occur in both the up and downstream flows. 

These simulations suggest that there are (at least!) two faults with our 1D model. Firstly, the treatment of mixing by a diffusive process is not ideal. Close to the boundary of the convective zone it may be a good approximation, but the further from the convective boundaries we go, the worse it gets. In the mid-regions of the intershell the fluid flow is advective and the transport of chemical species is rapid: there is not enough time for protons to burn in flight. A two-stream advective scheme may present a better approximation to the actual physical process. In addition, the use of velocities as calculated by mixing length theory may be significantly underestimating the speed of convective motions. 

\section{Conclusions}

We have modelled the ingestion of protons into the intershell convection zone in a low metallicity AGB star in a three dimensional hydrodynamics code. We find substantially higher hydrogen burning luminosities in our 3D model than in the input 1D evolutionary model. Rapid transport of hydrogen in downflowing plumes leads to vigorous hydrogen burning taking place in close proximity to the He-burning shell. We see no evidence for in-flight proton burning because the transport of protons from H-rich regions to the He-burning shell is so rapid. This rapid transport behaviour is not currently accounted for by 1D evolution codes, which use a diffusive approximation for their mixing of chemical species. We believe that the 1D codes could be substantially improved by using an advective, rather than a diffusive, mixing scheme for these episodes. Despite the extremely large hydrogen-burning luminosities found in our 3D simulations, we find no evidence for the convective region splitting into two zones. However, we caution that this may be because the simulations do not have sufficient spatial resolution to resolve the initial splitting or because they have not been evolved for long enough. However, it is also possible no splitting will occur in these models because the energy injection from proton burning occurs at the base of the convective region, rather than part-way through it. It should also be reiterated that higher resolution simulations are desirable to ensure that the boundary layer at the top of the intershell, and the entrainment of material in this layer, is properly resolved. Higher resolution simulations would also be expected to show more truly turbulent flow than those presented herein.

The ramifications of these simulations for stellar evolution remain to be determined and await the integration of these results into 1D evolution codes, which remain the only way we can investigate stellar behaviour over long timescales. This implementation is currently in progress but is only in its embryonic stages. Here we shall simply speculate on some possible consequences. If the current generation of 1D evolution codes is underestimating the amount of energy generation from hydrogen burning during proton ingestion events, then the depth of third dredge-up following one of these episodes is also likely underestimated. Carbon-rich material could be dredged to the surface earlier than current models predict. This could also mean that lower-mass stars, which may not have strong enough pulses to trigger third dredge-up, could in fact become carbon-rich. This could help to ameliorate the discrepancy between the observed carbon-rich to carbon-normal metal-poor star fraction and the predictions of population synthesis which require there to be more carbon-rich stars than theory currently predicts \citep[see][for further details of this problem]{2009A&A...508.1359I}. 

There will also be consequences for the nucleosynthesis of these events. The plentiful supply of protons to the \el{12}{C} rich intershell will generate a substantial quantity of \el{13}{C}. When this undergoes alpha-capture, a substantial quantity of neutrons will be liberated and this may give rise to an $s$ process \citep[see][for the case of proton ingestion during the core helium flash]{2010A&A...522L...6C}. In a 1D stellar model in which the splitting of the convective region was delayed (based upon their hydrodynamical simulations)  \citet{2011ApJ...727...89H} reported neutron densities of $10^{15}$ cm$^{-3}$ and significant production of the light $s$-process elements (Rb, Y, Sr and Zr). This situation was for a higher metallicity object undergoing a very late thermal pulse. Would similar nucleosynthesis take place in a low metallicity case like the one presented here? This is something that we will look at in future work.

\section{Acknowledgements}

We thank the referee, Casey Meakin, for his comments which have helped to improve this manuscript. We are very grateful to Lawrence Livermore National Laboratory for allowing us access to both {\sc djehuty} and the computers necessary to run it on. Without their support, this work would not have been possible. This work was partially supported by an LLNL Grand Challenge Grant for the study of convection in stars,  under the auspices of the US Department of Energy by Lawrence Livermore National Laboratory under contract DE-AC52-07NA2734. RJS is a Stromlo Fellow and acknowledges funding from the Australian Research Council Discovery Projects scheme (grant DP0879472) during his time at Monash. He is indebted to D. Arnett for illuminating discussions regarding turbulence and hydrodynamics in general. JCL acknowledges funding from the Australian Research Council Discovery Projects scheme (grants DP0877317 and DP1095368).

\bibliography{/Users/richardstancliffe/Work/NewBib}

\end{document}